\documentclass[apj,iop]{emulateapj}
\newcommand{\asec}{^{\prime\prime}}

\shorttitle{Parsec-scale mid-IR emission in NGC~424}
\shortauthors{H\"onig et al.}

\journalinfo{Accepted by The Astrophysical Journal}
\submitted{Submitted 2012 April 24; accepted 2012 June 20}

\begin{document}

\title{Parsec-scale dust emission from the polar region in the type 2 nucleus of NGC~424}

\author{S.~F. H\"onig,\altaffilmark{1} M. Kishimoto,\altaffilmark{2} R. Antonucci,\altaffilmark{1} A. Marconi,\altaffilmark{3} M.~A. Prieto,\altaffilmark{4} K. Tristram,\altaffilmark{2} G. Weigelt\altaffilmark{2}}% \altaffilmark{1}}
\altaffiltext{1}{University of California Santa Barbara, Department of Physics, Broida Hall, Santa Barbara, CA 93106-9530, USA; shoenig@physics.ucsb.edu}
\altaffiltext{2}{Max-Planck-Institut f\"ur Radioastronomie, Auf dem H\"ugel 69, 53121 Bonn, Germany}
\altaffiltext{3}{Dipartimento di Fisica e Astronomia, Universita' di Firenze, Italy}
\altaffiltext{4}{Instituto de Astrofsica de Canarias, La Laguna, Tenerife, Spain}

\begin{abstract}
Advancements in infrared (IR) interferometry open up the possibility to spatially resolve active galactic nuclei (AGN) on the parsec-scale level and study the circumnuclear dust distribution, commonly referred to as the ``dust torus'', that is held responsible for the type 1/type 2 dichotomy of AGN. We used the mid-IR beam combiner MIDI together with the 8\,m telescopes at the Very Large Telescope Interferometer (VLTI) to observe the nucleus of the Seyfert 2 galaxy NGC~424, achieving an almost complete coverage of the $uv$-plane accessible by the available telescope configurations. We detect extended mid-IR emission with a relatively baseline- and model-independent mid-IR half-light radius of $(2.0\pm0.2)\,\mathrm{pc}\,\times(1.5\pm0.3)\,\mathrm{pc}$ (averaged over the $8-13\,\micron$ wavelength range). The extended mid-IR source shows an increasing size with wavelength. These properties are in agreement with the idea of dust heated in thermal equilibrium with the AGN. The orientation of the major axis in position angle $\sim-27^\circ$ is closely aligned with the system axis as set by optical polarization observations. Torus models typically favor extension along the mid-plane at mid-IR wavelengths instead. Therefore, we conclude that the majority of the pc-scale mid-IR emission ($\ga$60\%) in this type 2 AGN originates from optically-thin dust in the polar region of the AGN, a scenario consistent with the near- to far-IR SED. We suggest that a radiatively-driven dusty wind, possibly launched in a puffed-up region of the inner hot part of the torus, is responsible for the polar dust. In this picture, the torus dominates the near-IR emission up to about $5\,\micron$, while the polar dust is the main contributor to the mid-IR flux. Our results of NGC~424 are consistent with recent observations of the AGN in the Circinus galaxy and resemble large-scale characteristics of other objects. If our results reflect a general property of the AGN population, the current paradigm for interpreting and modeling the IR emission of AGN have to be revised.
\end{abstract}



\section{Introduction}

\setcounter{footnote}{0}

The dusty environment around supermassive black holes in active galactic nuclei (AGN) came in reach of direct observations using the new high spatial resolution capabilities of long-baseline infrared (IR) interferometry. Within the last years several observing campaigns revealed emission sources in the nuclear region that shows clear characteristics of AGN-heated dust on the sub-parsec to parsec scale in the near- and mid-IR \citep[e.g.][]{Jaf04,Tri07,Bec08,Kis09a,Rab09,Tri09,Bur09}. The two prototypical type 2 AGN in NGC~1068 and the Circinus galaxy have been studied in detail using the mid-IR beam combiner MIDI at the Very Large Telescope Interferometer (VLTI) and pairwise combinations of the four 8\,m-telescopes. On these rather long baselines both objects were highly resolved with rather low visibilities that have been interpreted as the reemission signatures of the circumnuclear obscuring region, commonly dubbed ``dust torus''. This geometrically-thick torus is an essential part of the unification scheme of AGN and explains the difference of type 1 and type 2 AGN by angle-dependent obscuration. Recently, IR interferometry has shown that the spectral and spatial characteristics of the mid-IR emission sources (spectral slope and apparent size) in type 1 AGN are tightly correlated with the brightness profile and radial distribution of the dust \citep{Kis11b}.

One problem with these previous observations is that they either miss most of the single-aperture flux, which has been resolved out (NGC~1068 and Circinus), or that they do not have the position-angle coverage necessary to constrain the shape/orientation of the nuclear emission source. This structural information is an important input for torus models that simulate IR images and predict certain position-angle dependent spectral and spatial features \citep{Sch08,Hon10b}. 

In this paper, we present mid-IR interferometry observations of the Seyfert 2 galaxy NGC~424. These observations cover almost the entire $uv$-plane accessible with the VLTI UTs. NGC~424 is a SB0/a galaxy at a distance of 45.7\,Mpc. The galactic disk is inclined at about 70$^\circ$ based on an axis ratio of 0.36 \citep{Kir90}. It hosts an AGN that is classified as an optical type 2 AGN with polarized broad lines \citep[i.e., a hidden type 1 AGN;][]{Ver10} based on spectro-polarimetric observations \citep{Mor00}. The gas column towards the nucleus is quite large: \citet{Col00} report that the Hydrogen column density $N_H$ is probably in the Compton-thick regime ($\ga2\times10^{24}\,\mathrm{cm}^{-2}$), while \citet{Lam11} find less, but still significant, X-ray obscuration ($N_H \sim 1.7\times 10^{23}\,\mathrm{cm}^{-2}$). Both values are consistent with a heavily obscured AGN, and most of the obscuring column is probably intrinsic to the AGN, in spite of the considerable inclination of the host galaxy. Based on the velocity dispersion of the bulge, \citet{Bia07} estimate the mass of the central black hole as $\log M_\mathrm{BH} = 7.78$, leading to an Eddington ratio of $l_\mathrm{Edd} = 0.13$, which is typical for a Seyfert galaxy. The nucleus is radio-quiet and unresolved in radio \citep{Mun00}.

In Sect.~\ref{sec:obs}, we present the interferometric data and describe our data reduction strategy. For interpretation we also compiled a high spatial resolution IR SED of single-telescope data to complement the interferometry. In Sect.~\ref{sec:res}, wavelength- and position-angle-dependent sizes are extracted from the interferometric observations and the IR SED is analyzed. Both interferometry and photometry data are simultaneously modeled in Sect.~\ref{sec:model}. The results are discussed in Sect.~\ref{sec:disc} and put into context of AGN unification. Finally, we summarize our findings in Sect.~\ref{sec:summary}.

\section{Observations and data reduction}\label{sec:obs}

\subsection{Interferometry}\label{sec:obs:interferometry}

\begin{figure}
\begin{center}
\epsscale{1.2}
%\epsscale{0.8}
\plotone{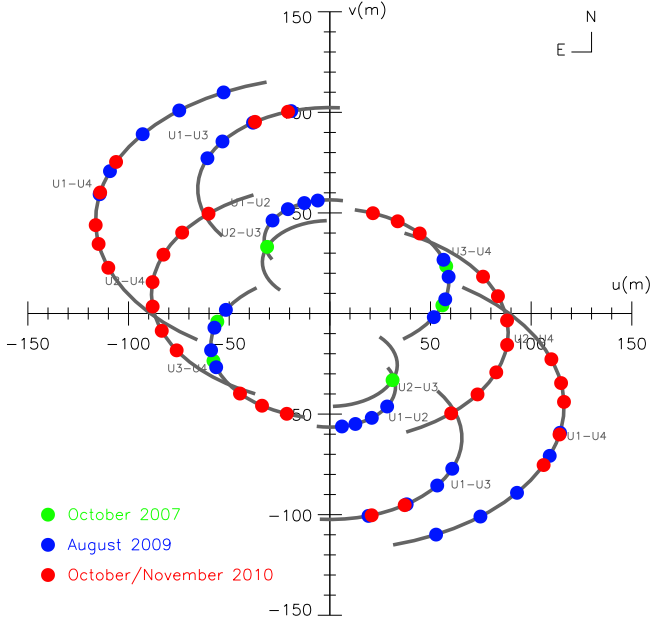}
\caption{$uv$-plane coverage (on-sky orientation) of the mid-IR interferometric observations of NGC~424. North is up, east is left.\label{fig:uv}}
\end{center}
\end{figure}
 
The interferometric observations have been carried out during three different campaigns in 2007, 2009, and 2010 using the mid-IR beam combiner \textit{MIDI} at the Very Large Telescope Interferometer (VLTI) in Chile. This instrument allows for spectro-interferometric observations in the mid-IR waveband from $8-13\,\micron$. We used several different combinations of the four 8-m telescopes to achieve various position angles and projected baseline points to fill the accessible $uv$-plane of NGC~424. In Fig.~\ref{fig:uv} we show the $uv$-plane coverage of our observations. The $uv$-plane is shown as projected on the sky to allow for an easy comparison with position angles. While the shorter baselines of up to 60\,m cover all quadrants, the longer baselines are oriented preferentially in NE/SW direction, in line with the location of the UT telescopes at the observatory. As illustrated, our observations can be considered complete for this target considering the achievable coverage using the VLTI UTs.

The data have been extracted by the standard MIA+EWS package developed for MIDI. Reduction and calibration were performed using the SNR maximization method as described in \citet{Kis11b}. Each individual fringe-track data set and a dedicated interferometric calibrator star have been reduced together to determine the required smoothing of the calibrator data to achieve the best signal-to-noise ratio of the correlated fluxes. During all three observing campaigns, we observed the same calibrators near NGC~424 (HD9362 and HD8498) for consistency, except for one set in 2007 where the only available calibrator was observed at significant spatial and temporal distance from the science target (HD219449). Cross-calibration showed that HD9362 and HD8498 remained essentially constant in flux and size over the observed time span. 

When two or three science fringe track sets have been observed together (i.e. within a 15\,min interval), they were averaged after data reduction with weights according to their individual SNR. This led to 3 independent correlated flux measurements in 2007, 17 in 2009, and 27 in 2010 as listed in Table~\ref{tab:obs}. The SNR values of all data sets range from 3.0 to 7.2 with a median of 4.3. This is slightly higher than for a typical AGN \citep[see][]{Kis11b} and can be explained by the mid-IR brightness of the source.

\begin{table}
\begin{center}
\caption{Observation log and data properties}\label{tab:obs}
\begin{tabular}{c c c c c c}
\hline\hline
Observing Date                & telescopes       & PBL     & PA      & calib. & max. \\ 
(UT)                                  &                         & (m)     & (deg)   &                   & SNR  \\ \tableline
2007--10--24 01:53      & U3--U4            & ~56.1 & ~94.2 & HD9362     & 4.1 \\ % 1
2007--10--24 03:57      & U3--U4            & ~63.0 & 112.3 & HD9362     & 6.2 \\ % 3
2007--10--25 04:00      & U2--U3            & ~45.5 & ~43.3 & HD219449 & 3.5 \\ \tableline % 1
2009--08--02 06:37      & U3--U4            & ~57.7 & ~88.0 & HD8498     & 4.4 \\ % 2
2009--08--02 07:37      & U3--U4            & ~57.8 & ~97.1 & HD8498     & 5.1 \\ % 2
2009--08--02 08:48      & U3--U4            & ~61.8 & 107.4 & HD9362     & 6.0 \\ % 1
2009--08--02 09:42      & U3--U4            & ~62.4 & 115.5 & HD9362     & 5.0 \\ % 2
2009--08--03 06:30      & U1--U2            & ~56.5 & ~~6.3 & HD8498     & 5.4 \\ % 2
2009--08--03 07:19      & U1--U2            & ~56.3 & ~13.1 & HD8498     & 6.5 \\
2009--08--03 08:26      & U1--U2            & ~55.9 & ~22.0 & HD9362     & 6.8 \\
2009--08--03 09:50      & U1--U2            & ~54.2 & ~31.7 & HD9362     & 7.2 \\
2009--08--04 06:23      & U1--U3            & 102.4 & ~11.0 & HD8498     & 3.0 \\
2009--08--04 07:37      & U1--U3            & 102.2 & ~22.0 & HD9362     & 3.0 \\
2009--08--04 08:52      & U1--U3            & 100.7 & ~32.3 & HD9362     & 3.2 \\
2009--08--04 09:46      & U1--U3            & ~98.2 & ~38.3 & HD9362     & 3.4 \\
2009--08--05 05:39      & U1--U4            & 122.0 & ~25.7 & HD8498     & 4.0 \\
2009--08--05 06:31      & U1--U4            & 125.7 & ~36.5 & HD8498     & 3.8 \\
2009--08--05 07:23      & U1--U4            & 128.8 & ~46.2 & HD8498     & 3.9 \\
2009--08--05 08:30      & U1--U4            & 130.1 & ~57.2 & HD9362     & 4.1 \\ % 1
2009--08--05 09:08      & U1--U4            & 128.7 & ~62.7 & HD9362     & 4.2 \\ \tableline % 1
2010--10--19 01:31      & U1--U3            & 102.4 & ~11.9 & HD9362     & 3.6 \\ % 1
2010--10--19 02:35      & U1--U3            & 102.3 & ~21.4 & HD9362     & 3.5 \\ % 2
2010--10--19 06:11      & U3--U4            & ~59.8 & 131.9 & HD8498     & 4.4 \\ % 2
2010--10--19 07:08      & U3--U4            & ~56.9 & 143.9 & HD8498     & 3.9 \\ % 2
2010--10--19 08:02      & U3--U4            & ~54.2 & 157.1 & HD8498     & 3.8 \\ % 2
2010--10--20 01:06      & U2--U4            & ~78.1 & ~50.5 & HD8498     & 4.8 \\ % 2
2010--10--20 01:58      & U2--U4            & ~83.7 & ~61.3 & HD9362     & 3.8 \\ % 2
2010--10--20 02:50      & U2--U4            & ~87.7 & ~70.5 & HD9362     & 4.4 \\ % 2
2010--10--20 03:49      & U2--U4            & ~89.4 & ~79.9 & HD8498     & 4.3 \\ % 3
2010--10--20 04:40      & U2--U4            & ~88.0 & ~87.8 & HD9362     & 4.3 \\ % 2
2010--10--20 05:32      & U2--U4            & ~83.9 & ~95.9 & HD8498     & 4.5 \\ % 2
2010--10--20 06:17      & U2--U4            & ~78.2 & 103.7 & HD9362     & 3.8 \\ % 1
2010--11--28 00:47      & U1--U4            & 130.2 & ~55.4 & HD8498     & 4.7 \\ % 2
2010--11--28 01:38      & U1--U4            & 128.7 & ~62.9 & HD8498     & 4.8 \\ % 1
2010--11--28 02:30      & U1--U4            & 123.6 & ~70.0 & HD8498     & 4.8 \\ % 1
2010--11--28 03:00      & U1--U4            & 119.0 & ~73.9 & HD9362     & 4.6 \\ % 1
2010--11--28 03:39      & U1--U4            & 111.2 & ~79.0 & HD9362     & 4.4 \\ \tableline  % 1
\end{tabular}
\end{center}
\end{table}

\begin{figure}
\begin{center}
\epsscale{1.2}
%\epsscale{0.8}
\plotone{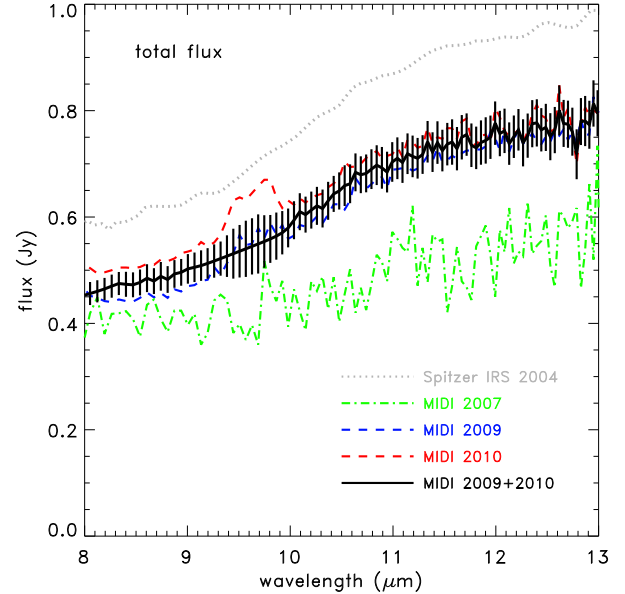}
\caption{Total $8-13\,\micron$ flux spectrum of NGC~424 as measured with MIDI. The red-dashed line represents the data from 2010, the blue-dashed line from 2009, and the green-dash-dotted line shows the 2007 data. The black-solid line is the error-weighted average of the 2009 and 2010 data and used as the reference for the interferometric data. The spectral feature around 9.6\,$\micron$ in the 2009 and 2019 data is a result of non-perfect subtraction of the atmospheric ozone feature.}\label{fig:tflux}
\end{center}
\end{figure}
 
We extracted the total fluxes from all of the photometry data that has been taken together with the fringe tracks. As sky reference we used the background flux from a 5 pixel offset position to the peak of the target flux. Calibration was done using the same calibrator star as used for the interferometric data. All data sets of each year have been averaged to obtain a reference total flux of NGC~424 in the $8-13\,\micron$ range for that year. The 2007, 2009, and 2010 spectra are shown in Fig.~\ref{fig:tflux}. While the 2009 and 2010 fluxes are consistent, the 2007 spectrum is systematically lower by a factor of 0.8 mostly independent of wavelength. 20--30\% variability over a span of 2 years may be possible, but given the bin-to-bin (systematic) scatter in the 2007 observations, it is more likely an observational inaccuracy. Indeed, the 2007 fluxes are based on only 3 photometry data sets while the 2009 and 2010 spectra are averages over 23 and 22 sets, respectively. For a good total flux reference, we calculate a ``master'' total flux spectrum, $F_\mathrm{tot}$, from the 2009 and 2010 data by averaging both sets and spline-fitting the atmospheric ozone feature at around 9.6\,$\micron$. This spectrum will be used as the reference for all correlated fluxes $F_{\mathrm{corr};i}$. The visibilities that we will use in the data analysis are then computed as $V_i=F_{\mathrm{corr};i}/F_\mathrm{tot}$, and visibility errors are propagated based on the total and correlated flux uncertainties. As described in \citet{Kis11b}, these uncertainties contain the fluctuations of the individual correlated and total flux frames, the differences to the mean when SNR-weighting several data sets, and a 5\% systematic component resulting from the analysis of low-SNR data \citep[see Appendix Fig. A.5 in][]{Kis11b}.

\subsection{Photometry}\label{sec:obs:photometry}

\begin{table}
\begin{center}
\caption{Nuclear IR photometry of NGC~424 extracted from archival VLT/ISAAC images.}\label{tab:irsed}
\begin{tabular}{c c c c c}
\tableline\tableline
filter & $\lambda$   & flux & calibrator & obs. date \\
         & ($\micron$) & (Jy)   &                 & (UT) \\ \tableline
Js      & 1.24             & 0.0043$\pm$0.0003 & S754--C & 2002--11--14 01:01 \\
H      & 1.65              & 0.0177$\pm$0.0007 & S754--C & 2002--11--14 01:05 \\
Ks     & 2.16             & 0.0445$\pm$0.0011 & S754--C & 2002--11--14 01:10 \\
L       & 3.78             & 0.2132$\pm$0.0046 & HD2811  & 2003--08--17 07:12 \\
M      & 4.66             & 0.3038$\pm$0.0294 & HD2811  & 2003--08--17 07:48 \\ \tableline
\end{tabular}
\tablecomments{The fractional contributions of the nuclear flux to the total flux in the central 1$\farcs$0 are 0.36, ($Js$-band), 0.57 ($H$-band), and 0.65 ($Ks$-band) based on the nuclear photometry after 2-D host subtraction with a Sersic profile in each band.}
\end{center}
\end{table}
 
In addition to our MIDI interferometry, we collected photometric data mainly from the ESO archive and literature to build a generic IR SED of the nuclear emission of NGC~424. The ESO data archive holds near-IR imaging data of NGC~424 in the $Js$-, $H$-, $Ks$-, $L$- and $M$-bands observed with ISAAC. We reduced the data using standard procedures. For the three near-IR bands, we first subtracted the host galaxy from the nucleus by fitting a Sersic profile to the extended ($>1\asec$) component (i.e. masking the nuclear region for the fit) before extracting nuclear fluxes. The resulting fluxes are shown in Table.~\ref{tab:irsed}. 

It is important to point out that the $L$- and $M$-band data from the 8m telescope with a (seeing-limited) spatial resolution of about 0$\farcs$4 are consistent, within error, with the $3.6\,\micron$ and $4.5\,\micron$ Spitzer data with a spatial resolution of about $1\farcs2$ reported by \citet{Gal10}. This can be taken as evidence that the IR emission in the central arcsecond is dominated by emission from the AGN without significant enhancement by starformation in the nuclear region. Indeed, the Spitzer IRS spectrum of the nuclear $\sim4\asec$ shown in \citet{Gal10} displays PAH features with only small equivalent widths, the largest being the 7.7\,$\micron$ feature but no line visible at 11.3\,$\micron$ \citep[][; AGN fraction $>0.9$ using the authors' relation between AGN contribution, IRAC and IRAS photometry]{Wu10}. 

\section{Results}\label{sec:res}

\subsection{Broad-band infrared SED}\label{sec:irsed}

\begin{figure}
\begin{center}
\epsscale{1.2}
%\epsscale{0.8}
\plotone{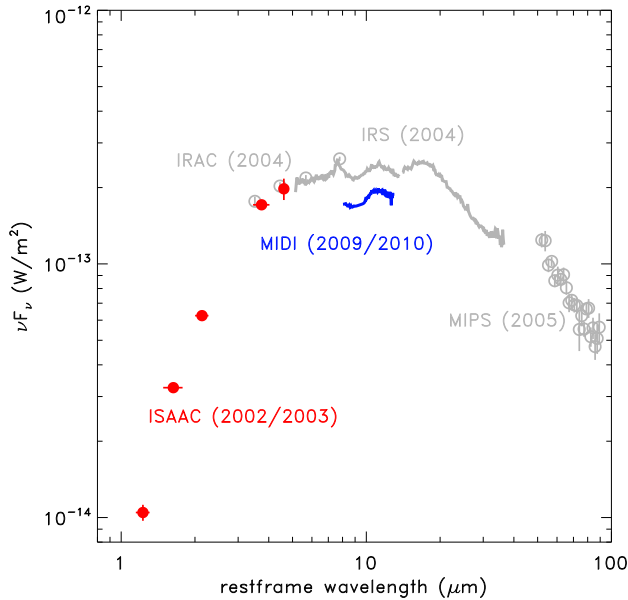}
\caption{$1-90\,\micron$ broad-band SED of NGC~424. The gray lines (IRS) and open circles (IRAC and MIPS) show the Spitzer data as presented by \citet{Gal10} with a spatial resolution of $>1.2\,\asec$ on the shorter and $>10\,\asec$ on the longer wavelength side. The red-filled circles represent the VLT/ISAAC data with a spatial resolution of $\sim0\farcs4$ while the blue line is our VLTI/MIDI total flux reference extracted from a window of about $0\farcs5 \times 0\farcs7$. The years in which the observations were taken are indicated.}\label{fig:irsed}
\end{center}
\end{figure}
 
In Fig.~\ref{fig:irsed} we show a broad-band IR SED from about $1\,\micron$ to $90\,\micron$ combining the Spitzer data as shown in \citet{Gal10} with the 8m-telescope data presented in this work. The plot illustrates the IR bump typical of dust emission and seen in all radio-quiet AGN. According to the unification scheme, it is caused by circumnuclear dust in the equatorial region of the AGN that re-emits the absorbed ``big blue bump'' (BBB) UV/optical radiation (presumed to originate from the accretion disk). One of the interesting properties of this target is that we see the IR SED rising at short wavelengths, flattening in the $8-20\,\micron$ region, and then falling off towards 100\,$\micron$, despite the comparable large intrinsic apertures involved. The resolution of the MIPS data at 70\,$\micron$ corresponds to $\sim$4.6\,kpc. Yet the SED is more or less a downward continuation of the data at shorter wavelengths and much higher angular resolution (at least to first order; see below). Indeed, the whole SED from the near- to far-IR connects smoothly despite the different instruments and resolutions involved. We interpret this as evidence that the unresolved IR nucleus dominates all of the IR emission without any significant contribution from other sources, e.g. circumnuclear starforming regions\footnote{If there were significant starformation within the photometric aperture, we would expect both strong PAH emission features and a bump in the far-IR. Neither is seen here though.}, and that the AGN is the only source of energy that produces the IR emission in the nuclear region.

There are two notable exceptions from the smooth connection of all SED data. The first one is a slight offset of the MIPS data with respect to the longest wavelengths in the IRS data. This may be an indication that there exists an additional, cool IR component contributing to longer wavelengths. If this were true, then we would expect that the slope changes from the downward tail of the IRS data through the MIPS data. However, this is not the case. In fact the data seem to be just offset by a factor of 1.3, which could be a sign of an issue with flux calibration in the far-IR, but we cannot exclude the possibility of slightly extended emission between the smaller and larger apertures. Anyway, the flux is strongly and monotonically decreasing toward the far-IR.

The second discontinuity in the SED is more interesting. While the ISAAC $L$- and $M$-band measurements are well in agreement with the IRAC data and smoothly connect to the IRS short wavelengths, the MIDI total flux spectrum is significantly lower than the IRS data. The difference of a factor of 1.3--1.4 is difficult to be attributed to calibration problems, since both the IRS and MIDI calibration seems to be consistent in itself (for MIDI see Sec.~\ref{sec:obs:interferometry}). Under the assumption that there are no unknown systematic problems with the MIDI data, we can imagine two possibilities for this discrepancy: first, there may be extended warm emission outside of the MIDI aperture of $\sim0\farcs6$ or 130\,pc. The dust cannot be too hot ($>$800\,K) or too cold ($<$200\,K), however, because we do not see a disagreement between small and large apertures in the near-IR and also no bump in the IRS spectrum beyond 20\,$\micron$. Moreover, the SED in the difference spectrum is flat, so that we can estimate a temperature of the possible extended emission of$\sim300-400\,\mathrm{K}$. When using the $r_\mathrm{sub}-L_\mathrm{12\,\micron}$-relation \citep{Hon10a} to estimate the dust sublimation radius for this type 2 source, we obtain about 0.065\,pc (with significant error bars). Given the constraints from the MIDI and Spitzer aperture, the emission responsible for the mid-IR flux difference would then be located at a distance $>2000\,r_\mathrm{sub}$. When assuming a standard ISM dust composition, we would expect an equilibrium temperature with the AGN radiation of $<$100\,K at that distance, which contradicts the requirements based on the SED shape as just discussed. Finally, an additional heating source that provides energy for the dust in a very narrow wavelength range is also difficult to imagine, in particular when required not to leave any other traces. We repeat that starformation is a very unlikely candidate because of the almost complete lack of PAH features (except a small feature at 7.7\,$\micron$) and the lack of a far-IR bump.

\begin{figure}
\begin{center}
\epsscale{1.2}
%\epsscale{0.8}
\plotone{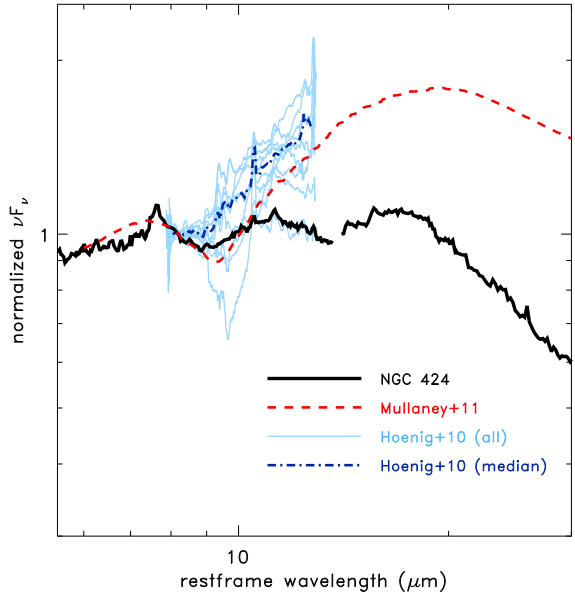}
\caption{Comparison of the 8\,$\micron$ normalized mid-IR SED of NGC~424 (black solid line) to AGN templates. The red-dashed line is an AGN template from \citet{Mul11} combining type 1 and type 2 AGN. The light-blue lines are 10 nearby AGN from \citet{Hon10a} observed with VLT/VISIR. These data have an intrinsic resolution between 30\,pc and 80\,pc at 8\,$\micron$ to approximately match the resolution of NGC~424 (55\,pc). The dark-blue dash-dotted line is the median spectrum of these AGN.}\label{fig:sedcom}
\end{center}
\end{figure}

The second possibility is mid-IR variability. In Fig.~\ref{fig:irsed} we indicated the years in which each of the sets of observations were taken. Interestingly the MIDI data is separated by about 4 years from the other observations. We recently discussed how IR variability of a dusty structure compares at different wavelengths \citep{Hon11b}. The principle of IR variability is simple: if the incident radiation onto the dust (here: the BBB radiation) changes, the dust temperature at a given distance from the source changes as well and, consequently, the flux varies. The main difference between near- and mid-IR wavelengths is the size and distance of the emitting regions. The near-IR emission is usually confined to a very narrow region close to the sublimation radius. Therefore the light-crossing time of this region for any variability signal is short and the initial signal is not significantly diluted (= comparable amplitudes of initial BBB signal and near-IR signal). The mid-IR emission is more complex. It is a combination of the Rayleigh-Jeans tail of hotter dust and the peak in emission from cooler dust. Moreover, the radial distribution of the dust also matters. Overall the mid-IR emission originates from a relatively large range of distances from the AGN, so that the variability signal is smeared out and the amplitudes are smaller, unless the initial signal has a very long duration \citep[e.g. a long-term break or jump in the AGN luminosity; see Fig.~2 in][for an example]{Hon11b}. Therefore, mid-IR variability is a viable possibility if both (1) the amplitude of variability and (2) the size of the region where this variability originates from are physically plausible. \citet{Gla04} reported near-IR variability in the $K$-band: the host-uncorrected $K$-band magnitude showed peak-to-valley variation of 0.34 mag over several years. Comparing the uncorrected $K$-band magnitude with our nuclear $K$-band flux reveals that about 70\% of the $K$-band flux of \citet{Gla04} originates from the host. The corrected $K$-band variability is, therefore, a factor of $\sim1.8$ using our nuclear $K$-band measurement. This near-IR amplitude is sufficient to explain a factor of 1.4 variability in the mid-IR. The second necessary condition related to the size of the mid-IR emission region can be constrained by the elapsed time between the epochs of the mid-IR observations: Within the 4 years, the light propagated about 1.2 pc, so that in order to obtain the potential decrease in mid-IR flux by a factor of 1.4, the bulk of the mid-IR emission must come from within this radius. We can test this possibility with our interferometric data (see below).

The SED of NGC~424 is relatively blue when compared to other type 2 or even type 1 AGN. In Fig.~\ref{fig:sedcom} we show the mid-IR emission of NGC~424 (black solid line) and two different AGN template SEDs. The red-dashed line is an AGN template from \citet{Mul11} where Spitzer IRS data of 25 AGN were corrected for galactic/starformation and averaged without distinguishing among types. We already showed that the Spitzer IRS flux of NGC~424 is consistent with the IR fluxes from ground-based 8\,m data. These data have higher angular resolution than Spitzer data. At 8\,$\micron$, the PSF in NGC~424 has a size of $\sim$55\,pc. To account for possible resolution effects in the \citet{Mul11} template, we used the 10 AGN from \citet[][light-blue lines]{Hon10a} with similar resolution as NGC~424 (range of 30$-$80\,pc at 8\,$\micron$) and calculated a median template (dark-blue dash-dotted line). NGC~424 shows a bluer SED than the combined SED. The spectral index $\alpha$ (defined as $F_\nu \propto \nu^\alpha$) is $\alpha\sim-2.1$ for type 2s and $\alpha\sim-1.7$ for type 1s, approximately in agreement in \citet{Hon10a} and \citet{Mul11}. NGC~424 has a mid-IR spectral index of $\alpha\sim-1.1$, making it the bluest type 2 compared to the other objects in \citet{Hon10a} with only one type 1 AGN showing a bluer mid-IR spectrum. This points toward a lack of cool dust relative to the amount of hot dust in NGC~424.

\subsection{Interferometry data analysis}

\begin{figure*}
\begin{center}
\epsscale{1.2}
%\epsscale{0.95}
\plotone{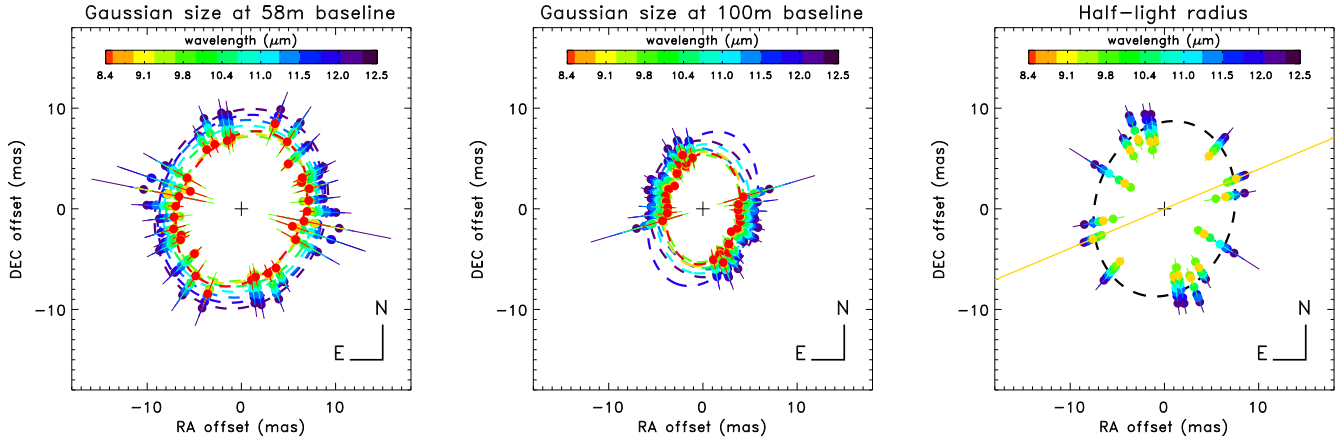}
\caption{Sizes of the mid-IR emission of NGC~424 for different wavelengths from 8.4\,$\micron$ (red circles) to 12.5\,$\micron$ (blue circles), determined from our MIDI interferometry data using different methods. \textit{Left:} Gaussian HWHM sizes (in mas) at a fixed baseline of 58$\pm$3\,m for a range of position angles (PAs). \textit{Center:} Gaussian HWHM sizes (in mas) at a fixed baseline of 100$\pm$3\,m for a range of PAs. Where no individual data were available, data pairs of similar PA have been interpolated or extrapolated. The colored-dashed lines are elliptical fits to the data at the corresponding wavelength bin. \textit{Right:} Half-light radius for a range of PAs. The black-dashed line is an elliptical fit to the wavelength-averaged data.}\label{fig:res:allsizes}
\end{center}
\end{figure*}

The visibilities and correlated fluxes of all reduced 37 datasets are shown in the online-only Fig.~\ref{fig:app:intf}. These data will be analyzed in the following.

\subsubsection{The size of the mid-IR source in NGC~424}

\begin{table*}
\begin{center}
\caption{Geometric fits to the visibilities of NGC~424.\label{tab:res:ellfit}}
\begin{tabular}{c c c c c c c c c c}
\tableline\tableline
fit type                     & wavelength  & \multicolumn{3}{c}{semi-major axis $a$}      &  & \multicolumn{3}{c}{semi-minor axis $b$}                                & position angle \\ \cline{3-5}\cline{7-9}
                            & ($\micron$) &       (mas)        &         (pc)        & ($r_\mathrm{in}$) & &       (mas)        &         (pc)        & ($r_\mathrm{in}$) & (deg)                \\ \tableline
Gaussian             &  9.1 & $ 8.0 \pm  1.0$ & $ 1.7 \pm  0.2$ & $26 \pm  3$ & & $ 6.5 \pm  0.7$ & $ 1.4 \pm  0.2$ & $21 \pm  2$ & $-40 \pm  23$ \\
(PBL 58\,m)         & 11.0 & $ 8.6 \pm  1.2$ & $ 1.9 \pm  0.3$ & $28 \pm  3$ & & $ 7.9 \pm  1.0$ & $ 1.7 \pm  0.2$ & $25 \pm  3$ & \textit{...} \\
                           & 12.5 & $10.2 \pm  1.7$ & $ 2.2 \pm  0.4$ & $33 \pm  5$ & & $ 8.9 \pm  1.8$ & $ 1.9 \pm  0.4$ & $29 \pm  5$ & \textit{...} \\
                           & 8.4--12.5 & $ 8.7 \pm  0.9$ & $ 1.9 \pm  0.2$ & $30 \pm  3$ & & $ 7.6 \pm  1.0$ & $ 1.7 \pm  0.2$ & $26 \pm  3$ & $-40 \pm   8$ \\ \tableline
Gaussian              &  9.1 & $ 5.7 \pm  1.5$ & $ 1.2 \pm  0.3$ & $18 \pm  4$ & & $ 4.0 \pm  0.3$ & $ 0.9 \pm  0.1$ & $13 \pm  1$ & $-12 \pm  26$ \\
(PBL 100\,m)        & 11.0 & $ 6.5 \pm  1.6$ & $ 1.4 \pm  0.4$ & $21 \pm  5$ & & $ 4.9 \pm  0.6$ & $ 1.1 \pm  0.1$ & $16 \pm  1$ &\textit{...} \\
                            & 12.5 & $ 6.8 \pm  1.1$ & $ 1.5 \pm  0.2$ & $22 \pm  3$ & & $ 5.7 \pm  1.3$ & $ 1.2 \pm  0.3$ & $18 \pm  4$ & \textit{...} \\
                            & 8.4--12.5 & $ 6.2 \pm  0.8$ & $ 1.4 \pm  0.2$ & $21 \pm  2$ & & $ 4.7 \pm  0.7$ & $ 1.0 \pm  0.2$ & $16 \pm  2$ & $ -5 \pm   11$ \\ \tableline
Half-light radius   &  9.1--12.5 & $ 9.1 \pm  0.9$ & $ 2.0 \pm  0.2$ & $ 31 \pm  3$ & & $ 7.0 \pm  1.1$ & $ 1.5 \pm  0.3$ & $ 24 \pm  4$ & $-21 \pm  3$ \\ \tableline
\end{tabular}
\end{center}
\end{table*}

\paragraph{A note on interferometric sizes} 

The correlated fluxes can be compared to the total fluxes to obtain visibilities that, in turn, can be converted into source sizes using a suitable model. AGN sources have often been modeled by means of a Gaussian flux distribution resulting in sizes based on the Gaussian FWHM or HWHM. However, it was shown recently that sizes inferred from single-Gaussian models can be quite misleading since the underlying brightness distribution barely resembles a Gaussian \citep{Kis11b}. In fact, these sizes can be considered rather a measure of the array resolution than the source size. This becomes even more of an issue when 2-dimensional sizes are determined from data where the baseline lengths change with position angle -- meaning the source shape is depending on the $uv$-coverage, a typical problem in contemporary IR interferometry owing to the limiting number of telescopes in the arrays. There are two ways to approximately overcome this problem. First, one chooses a fixed visibility, infer the baseline lengths/spatial wavelengths for all position angles at this visibility, and calculates a corresponding size. This has been done in \citet{Kis11b} in terms of the half-visibility radius $R_{V0.5}$ for a fixed visibility of 0.5 or related half-light radius $R_{1/2}$. Second, it is also possible to take a fixed projected baseline length (PBL) and infer the visibilities for all position angles at this PBL. The first method is more related to the intrinsic structure of the object while the second one reflects the shape that a telescope with an aperture the size of the PBL would see. Both methods, however, overcome the problem of inhomogeneous $uv$-coverage to some extent.

\paragraph{NGC~424 emission source}
 
\begin{figure}
\begin{center}
%\epsscale{0.8}
\epsscale{1.2}
\plotone{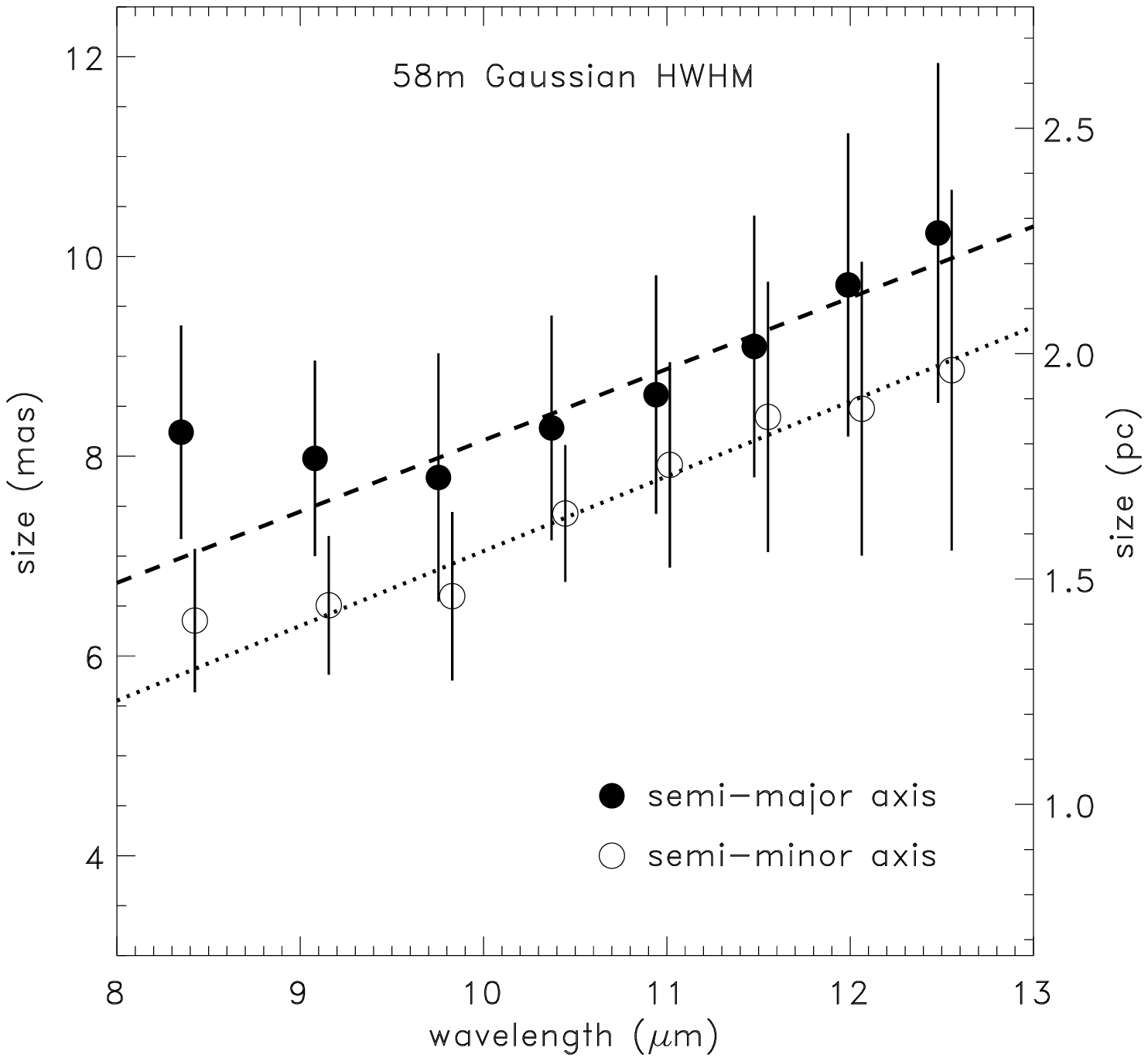}
\caption{Wavelength-dependence of the Gaussian HWHM size along the major axis (filled circles) and minor axis (open circles) at a fixed baseline length of 58\,m. Please note that the error bars are larger than the bin-to-bin variance because they are dominated by systematic errors in the visibility. The dashed and dotted lines are linear fits to the data at $>9\,\micron$ for the semi-major and semi-minor axis, respectively.}\label{fig:res:wlsize}
\end{center}
\end{figure}
 
In the left panel of Fig.~\ref{fig:res:allsizes}, we show the position-angle-dependent Gaussian HWHM sizes of the mid-IR emission source in NGC~424 at a fixed PBL of 58\,m as obtained from our data. Where no data were available in the range of 55--61\,m, we used pairs of data at similar position angles (PAs) and interpolated or extrapolated to 58\,m (including errors). \citep{Kis11b} showed that the brightness distributions in AGN follow simple power laws, so that we linearly inter-/extrapolated in log-space. Because of the resulting well-sampled PA coverage, some of the source characteristics can be easily inferred. First, the object is smallest at the shortest wavelength bin and becomes larger with wavelength (see also Fig.~\ref{fig:res:wlsize}). This is qualitatively consistent with the idea that hotter dust (primarily emitting shorter wavelengths) is located closer to the AGN. Second, the mid-IR source is non-circular. While this is not too obvious in the 58\,m data, we also show Gaussian HWHM sizes at 100\,m in the middle panel of Fig. ~\ref{fig:res:allsizes}. The sizes in East/West direction are clearly smaller than those close to North/South for all wavelengths. It should be noted, however, that there are larger gaps in the PA coverage at 100\,m because of the limited $uv$-coverage by the UT telescopes (see Fig.~\ref{fig:uv}). In the right panel of Fig.~\ref{fig:res:allsizes}, we show the PA-dependence of the half-light radius in NGC~424. As previously noted, this measure is independent of the actual baseline length. However, data interpolation is reversed with respect to the fixed-baseline method and results in larger uncertainties of the spatial wavelength because the change in visibilities with baselines is small. In the right panel of Fig.~\ref{fig:res:allsizes}, we show all data pairs that allow us to establish a clear half-light radius. The problem is particularly relevant to the shorter wavelength data as can be seen by the sometimes erratic size changes at 9.1\,$\micron$ (orange circles). Taking all three panels together makes it clearer that the objects displays some degree of elongation.

To be more quantitative about the object's size and shape, we fitted ellipses to the different measurements. In the left and middle panel of Fig.~\ref{fig:res:allsizes} we show fits to each wavelength bin of the Gaussian sizes at 58\,m and 100\,m. Because of the higher uncertainties in the determination of the half-light radius, only the median visibility of all wavelengths was used for the fit in the right panel of Fig.~\ref{fig:res:allsizes}. In Table~\ref{tab:res:ellfit} some reference fitting results are listed. The size is typically between 0.9\,pc and 2.2\,pc, depending on baseline length, wavelength, and orientation. The half-light size of $(2.0\pm0.3)\,\mathrm{pc}\,\times(1.5\pm0.3)\,\mathrm{pc}$ would make a variability origin of the difference between the MIDI and Spitzer+ISAAC fluxes physically possible (see Sect.~\ref{sec:irsed}), but we note again that the difference spectrum is cool. Therefore, a final conclusion about the origin of the flux difference cannot be made. For the sizes please note that due to the $uv$-coverage, the largest (relative) uncertainties in the sizes are toward the semi-minor axis for the 58\,m fit and toward semi-major axis for the 100\,m fit. 

We present the sizes in observed (milliarcseconds) and physical (pc) units, as well as in intrinsic units, i.e. in units of the sublimation radius. The latter one is very useful when we want to compare properties in different objects with different luminosities \citep{Kis09b,Kis11b}. For that we have to establish an inner radius, $r_\mathrm{in}$ of the dust distribution (also referred to as sublimation radius). Usually this is inferred either from reverberation mapping or near-IR interferometry \citep[e.g.][]{Sug06,Kis09a,Kis11a}. Since we do not have either information in this type 2 AGN, we use an estimate based on the 12\,$\micron$ luminosity as discussed in \citet{Hon10a}. Taking the mean between Spitzer and MIDI total flux, we obtain $r_\mathrm{in} \sim 0.065\,\mathrm{pc}$ for a luminosity of $\nu L_\nu(12\,\micron) = 5.4\times10^{43}\,\mathrm{erg/s}$. 

The ellipse fits illustrate some key characteristics of the mid-IR emission source. They confirm the obvious trend that the source size increases with wavelength, as illustrated in Fig.~\ref{fig:res:wlsize}. Although the error bars are large (dominated by systematic errors in the visibility), a systematic trend of larger mid-IR emission sources for longer wavelengths is seen.  
%Moreover, it seems as if the size along the major axis $a$ increases more rapidly with wavelength than towards the minor axis $b$. The nominal ratios for 12.5--to--9.1\,$\micron$ size in $a$-direction are about 1.2  for both 58\,m and 100\,m, while they are about 1.4 in $b$-direction. Given the error bars of the individual observations, this difference might seem insignificant, however, these results are consistent at both PBLs and might indicate a real trend. 

\subsubsection{Position angle dependencies of the mid-IR source}\label{sec:res:orient}

\begin{figure}
\begin{center}
\epsscale{1.2}
%\epsscale{0.8}
\plotone{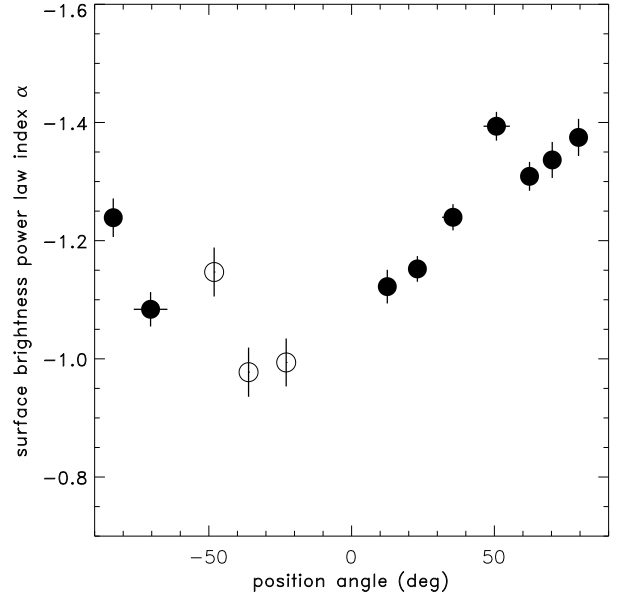}
\caption{Position-angle dependence of the mid-IR brightness distribution, parameterized by a surface density power law with index $\alpha$ and a fixed temperature law (see text for details). The filled circles mark PAs where the brightness distribution was modeled by two or three sets of interferometric observations at different baseline lengths, while the open circles denote PAs with just one baseline.}\label{fig:res:radbright}
\end{center}
\end{figure}
 
The elliptical fits also provide information of the direction of elongation. In Table~\ref{tab:res:ellfit} we list the position angle of the half-light radius for those fits that allow for such a constraint. While the formal fitting errors are considerable, all position angles systematically indicate an elongation toward the NW/SE direction. The fits that average over all wavelengths give approximately $-5^\circ < \mathrm{PA} < -40^\circ$. There may be a trend that the position angle changes slightly from the lower-resolution data at 58\,m to the 100\,m data. While such a trend would be reminiscent of interferometry data of the Circinus \citep{Tri12}, the uncertainty in the geometric determination of the PA do not allow us to formally establish such a skew (see also the PA for the half-light radius), in particular because the 100\,m data do not have any coverage in the direction of the major axis. Therefore, any position angle determined for the major axis in the 100\,m data must be considered uncertain.

We note that this elongation is not affected by the inhomogeneous $uv$-coverage, since this has been taken care of as discussed above. However, when deriving sizes for each wavelength, only the spatial information of the interferometric data is used. We will now include the spectral information in our analysis to test the result from the geometric model fitting. For that we use a simple model of dust emission to characterize the mid-IR emission at as many position angles as possible that are covered by our observations. For each PA, we simulate a brightness distribution of the form
\begin{equation}\label{eq:1}
S_\nu(r) = B_\nu(T(r)) \cdot f_0\left(\frac{r}{r_\mathrm{in}}\right)^\alpha
\end{equation}
where $r$ is the distance from the AGN, $r_\mathrm{in}$ the inner radius of the dust distribution, $f_0$ is the normalized surface density at the inner radius, $\alpha$ is the surface density power law index, and $B_\nu$ the black-body radiation with temperature $T(r)$ at distance $r$. In the following analysis we will fix $f_0=1$ for simplicity but use it in Sect.~\ref{sec:model} as a free parameter. For the temperature we assume that the dust is in local thermal equilibrium and parameterize the temperature gradient by $T(r) = T_\mathrm{sub} \cdot (r/r_\mathrm{in})^\beta$. We use a sublimation temperature $T_\mathrm{sub} = 1\,500\,\mathrm{K}$ and adopt $\beta=-0.36$ from the near-IR dust opacity model by \citet{Bar87}. The only remaining parameter is the power-law index $\alpha$ given the $r_\mathrm{rin}$ value estimated in Sect.~\ref{sec:irsed}. This optically-thin model describes the directly AGN-heated dust as we see it in the mid-IR, while any possible cooler dust is not covered. However, the far-IR SED does not suggest the presence of such a cool component (see Fig.~\ref{fig:irsed} and Sect.~\ref{sec:irsed}).

In Fig.~\ref{fig:res:radbright} we show the PA-dependence of the surface density distribution power law index $\alpha$. Where applicable we fitted the model to two or three data sets with different baseline length (filled circles). If no second set was available within 10$^\circ$ single baseline data was used (open circles). The data clearly show a change of $\alpha$ with PA. From about PA $-90^\circ$ to $-45^\circ$, the power law index becomes less negative. Toward larger positive PAs, $\alpha$ turns more negative with a peak in the range of $50-80^\circ$. This change can be either interpreted as a PA-dependence of the brightness distribution (i.e. traced back to a change in $\alpha$, $\beta$, or a combination of both), or a PA-dependence of $r_\mathrm{in}$, which is used for normalization. In either case it illustrates that, indeed, the mid-IR source changes structurally with position angle, and the shallowest $\alpha$ values (=least negative) in PA~$\sim-30^\circ$ corresponds to about the same direction as the major axis in the geometric fits.

\section{Simultaneously modeling the SED and interferometry of NGC~424}\label{sec:model}

\subsection{Model description}\label{mod:setup}

The analysis of the NGC~424 interferometric data presented so far was based on the observations with as little model-dependence as possible in order to extract basic properties of the mid-IR emission. However, only specific aspects of the data have been investigated. Now, we want to model the full data set of mid-IR interferometric observations together with the SED. This can be done, in principle, with advanced models of clumpy tori \citep[for such examples see][]{Hon06,Hon10a,Tri11}. On the other hand, the visibilities in NGC~424 do not go below 0.4 and models suggest that effects that clearly distinguish homogeneous models from clumpy ones will only appear when an object is better resolved \citep{Sch08,Hon10b}. This means the present observations are not sensitive to substructure but only to the overall distribution of the brightness. 

We, therefore, decided to use a rather simple physical model to reproduce the interferometric and photometric data simultaneously to avoid over-interpretation. The model follows \citet{Kis11b} and was successfully applied to type 1 AGN. The brightness distribution is defined by the dust surface density power law index $\alpha$ for the emitting dust and the temperature gradient $\beta$ of the dust (see Eq.~\ref{eq:1}), plus an additional hot ring component (see below). Here, $\beta$ is left as a free parameter and $T(r)$ starts at a maximum temperature $T_\mathrm{max}$ that we also fit for. The surface density $f_0$ at the inner radius will be used as a free parameter now (see Sect.~\ref{sec:res:orient}). We note that we do not need an ad-hoc assumption of $r_\mathrm{in}$ but can leave it as a free parameter as well. It sets the normalization of both the fluxes and the visibilities: For a given set of parameters, a larger $r_\mathrm{in}$ will result in higher fluxes (larger emitting surface) and lower visibilities (larger sizes). As already discussed, the mid-IR emission is elongated. Therefore, we describe the source shape, and as a consequence the brightness profile, as an ellipse with the ratio $s=a/b$ between the semi-major and semi-minor axis and orientation $\phi$ of the major axis as parameters\footnote{Since NGC~424 is a type 2 AGN, probably at considerable inclination, the meaning of a ring as description for a projected face-on torus as in a type 1 AGN becomes obsolete. The concept of an inner radius with inclusion of ellipticity, however, captures the fundamentals of dust emission and sublimation and axial symmetry.}.

\begin{table*}
\caption{Best-fit parameters for the modeling of the IR photometry and interferometry.\label{tab:mod:fit}}
\begin{center}
\begin{scriptsize}
\begin{tabular}{c c c c c c c c c c}
\tableline\tableline
model                  & $\alpha$ & $\beta$ & $\phi$ & $s=a/b$ & $T_\mathrm{max}$ & $f_0$ & $f_\mathrm{hot}$ & $r_\mathrm{in}$ & $\chi^2_r$\\
                            &        &        &  (deg)   &        &   (K)                &       &                          & (pc) & \\ \tableline
\multicolumn{10}{c}{\textbf{photometry + interferometry}} \\
hot ring + $T/d$-gradient                            & $-0.97\pm0.05$ & $0.31\pm0.03$ & $-27\pm2$ & $2.2\pm0.1$ & $570\pm63$ & $0.81\pm0.20$ & $0.05\pm0.02$ & $0.101\pm0.013$ & 0.86 \\ 
$T/d$-gradient (w/o $L$, $M$)                    & $-0.85\pm0.14$ & $0.32\pm0.03$ & $-27\pm1$ & $2.2\pm0.1$ & $662\pm81$ & $0.99\pm0.25$ & $\ldots$            & $0.066\pm0.013$ & 0.83 \\ 
$T/d$-gradient (MIDI only)                           & $-1.00\pm0.18$ & $0.31\pm0.04$ & $-27\pm2$ & $2.2\pm0.1$ & $713\pm101$ & $0.94\pm0.30$ & $\ldots$            & $0.063\pm0.016$ & 0.80 \\ 
\tableline
\multicolumn{10}{c}{\textbf{photometry only}} \\
%hot ring + $T/d$-gradient      & \multicolumn{8}{c}{unconstrained} \\
hot ring + $T/d$-gradient      &                      $>-0.5$ & $<0.36$ & $\ldots$ & $\ldots$ & $>900$ & $<0.3$ & $>0.1$ & $<0.05$ & \\ %new
$T/d$-gradient (w/o $L$, $M$) & $-1.02\pm0.25$ & $0.32\pm0.08$ & $\ldots$      & $\ldots$        & $545\pm154$ & unconstr. & $\ldots$          & unconstr. & 0.32 \\ % new
\tableline
\end{tabular}
\tablecomments{Model parameters: $\alpha\ldots$power law index of the surface density distribution; $\beta\ldots$temperature-gradient power law index; $\phi\ldots$position angle of the major axis (in degrees from N toward E); $s\ldots$ratio of major axis $a$ and minor axis $b$; $T_\mathrm{max}\ldots$maximum temperature of the dust distribution; $f_0\ldots$normalized surface density at the inner radius $r_\mathrm{in}$; $f_\mathrm{hot}\ldots$fractional contribution of hot-dust component; $r_\mathrm{in}\ldots$inner radius of the dust distribution}
\end{scriptsize}
\end{center}
\end{table*}
 
The hottest emission in this model originally designed for type 1 AGN is coming from a ring (in addition to the power law) at about the inner radius with temperature $T_\mathrm{in}=1\,400\,\mathrm{K}$ and with a contribution factor $f_\mathrm{hot}$. This hot ring has been held responsible for the observed near-IR bump in the SEDs of many type 1 AGN \citep[e.g.][]{Ede86,Kis11b,Mor11}. Since NGC~424 is a type 2 AGN, we will show model fits with and without the hot ring. The origin of this hot emission component is not fully understood. It may either be the result of a different dust size/composition in the hot region of the torus (silicates sublimate at cooler temperatures than graphite grains; small grains heat up to the sublimation temperature at larger distances), or coming from a graphite-dust component in the innermost part of the polar outflow region. We will discuss an addition possibility in Sect.~\ref{sec:para}. In any case the near-IR bump is typically not seen in type 2 AGN, indicating significant extinction towards this component. From the presence of the silicate features in \textit{emission}, we conclude that the dust column toward a hot dust component may be optically thin at $>8\,\micron$. From about $4-8\,\micron$ typical ISM dust extinction curves do not change significantly (but note that the dust may have a non-ISM composition; see Sect.~\ref{sec:para}). However, towards the near-IR $J$, $H$, and $K$ bands -- the peak regions of the hot dust component -- the extinction increases dramatically by a factor of $5-10$. To avoid these subtle problems, we ignore the shortest wavelengths in the fitting process.

\subsection{Data selection}\label{mod:data}

As discussed in the previous section, there may be several problems involved when trying to fit the near-IR photometric data, most notable the possible extinction. We, therefore, restrict our modeling efforts to the data $>3\,\micron$. In order to capture the downturn of the SED longward of 20\,$\micron$, but not overweighting the photometric data over the interferometric, we extract fluxes at 22\,$\micron$ and 32\,$\micron$ from the Spitzer data and treat them as photometric data points. In addition, to account for the offset of the MIDI data with respect to the rest of the SED (see Sect.~\ref{sec:irsed}) the photometry has been corrected by a factor of 1.3, so that the Spitzer IRS $8-13\,\micron$ matches the MIDI spectrum.

To speed up the fitting process and gain SNR, we resampled and binned the interferometric data. The $8-13\,\micron$ range will be represented by 9 wavelength bins per data set in the fitting process (see Fig.~\ref{fig:app:intf} in the appendix). While the 2009 and 2010 data are in general agreement for all baseline configurations that match between both epochs, two of the 2007 data sets show significant disagreement. In particular, the two U3-U4 sets with only moderately different PBL and PA (see Table.~\ref{tab:obs}) are inconsistent with each other. There are several technical difficulties involved when dealing with the 2007 data (e.g. the IRIS beam-tracking system of the VLTI did not work properly at that time on targets with $K$-magnitude $\sim10$, resulting in poor/unstable beam overlap) in addition to the low-quality total flux reference of 2007 (see Sect.~\ref{sec:obs:interferometry}), so that we will not include these data in the fitting process. However, the fits will be overplotted onto the 2007 data in Fig.~\ref{fig:app:intf} as a comparison.

Last, the visibilities and correlated fluxes show a distinct change of slope at the shortest wavelength bin at about 8.4\,$\micron$ with respect to the rest of the spectrum. There are two possibilities that might explain this feature in the data. First, the change of slope could be real. As a consequence the wavelength-dependent sizes would not become smaller with decreasing wavelength anymore. Indeed the 8.4\,$\micron$ sizes may be even systematically larger than for the neighboring 9.1\,$\micron$ bin (see Fig.~\ref{fig:res:wlsize}). While this is difficult to explain with in simple radiative equilibrium, it could be an effect of extinction/obscuration at shorter wavelengths. However, as discussed in the previous section, the overall increase of extinction toward shorter wavelengths in this range is rather small for any kind of dust (ISM, large grains, silicates, graphites). Another possibility for a real decrease in visibility is contamination by the 7.7\,$\micron$ PAH feature. Indeed, this feature is weakly present in the Spitzer IRS spectrum, and the wings may still be seen on the short-wavelength side of our MIDI total flux spectrum. If these wings are not present in the correlated fluxes, but only in the total flux, the visibility would be expected to be lower, as seen here. However, in this case the lower visibilities are not physically related to the parsec-scale dust distribution. Second, the lower 8.4\,$\micron$ visibilities/correlated fluxes may be an instrumental/calibration effect. Indeed the sensitivity of the MIDI detector is sharply changing at these short wavelengths, in addition to generally lower flux levels than at longer wavelengths and shorter coherence times. To avoid problems with these lower visibility at 8.4\,$\micron$ that are most likely not related to the parsec-scale dust, we will exclude it from the fitting process (i.e. using exclusively $9-13\,\micron$ data).

\subsection{Model results}

In Table~\ref{tab:mod:fit} we list the results for a number of different modeling approaches (models as described in Sect.~\ref{mod:setup}). The first model uses all the data as described in Sect.~\ref{mod:data} and fits the temperature/density gradient model plus the hot ring. The second fit excludes the $L$- and $M$-band photometry as well as the hot ring that is essentially only necessary to account for these wavelengths (see Sect.~\ref{mod:setup}). The resulting reduced $\chi^2_r$ of both models are equal, underlining this point. Finally we use only the MIDI data in the $9-13\,\micron$ range and model it with the temperature- and density- gradient (or $T/d$-gradient) model without hot ring, again with a $\chi^2_r\sim0.8$ as the previous fits. 

The resulting parameters point toward a surface density distribution with power-law index $\alpha\sim-0.9\ldots-1.0$. This result is consistent with the more model-independent determination of $\alpha$ illustrated in Fig.~\ref{fig:res:radbright}, taking into account that the more physical model accounts for the PA-dependence by ellipticity. The value of $\alpha$ is rather on the steeper side when compared to type 1 AGN \citep{Kis11b}, which is particularly interesting since torus models would predict generally redder colors for type 2s than type 1s because of obscuration effects. 

The orientation of the mid-IR emission is quite similar to what we got from the geometric fits (consistent within error bars). A PA of $-27^\circ \pm 2^\circ$ is within the range of PAs as listed in Table~\ref{tab:res:ellfit}. The $a/b$ ratio of 2.2 in the model illustrates the strong ellipticity of the mid-IR source. This ratio may seem larger than visible in Fig.~\ref{fig:res:allsizes}, but these perceived differences come, at least in part, from resolution effects (see also the stronger ellipticity for the 100\,m data than for the 58\,m data)\footnote{Similar to single-telescope data, the observed/measured object shape depends on the intrinsic shape convolved with a ``resolution kernel'' or PSF. The size of this kernel depends on the baseline lengths and has larger widths for smaller baseline lengths. This leads to ``smearing'' of the actual shape of the object. In an unresolved case, the observed shape would be the PSF of a point source.}.

\begin{figure}
\begin{center}
\epsscale{1.2}
%\epsscale{0.8}
\plotone{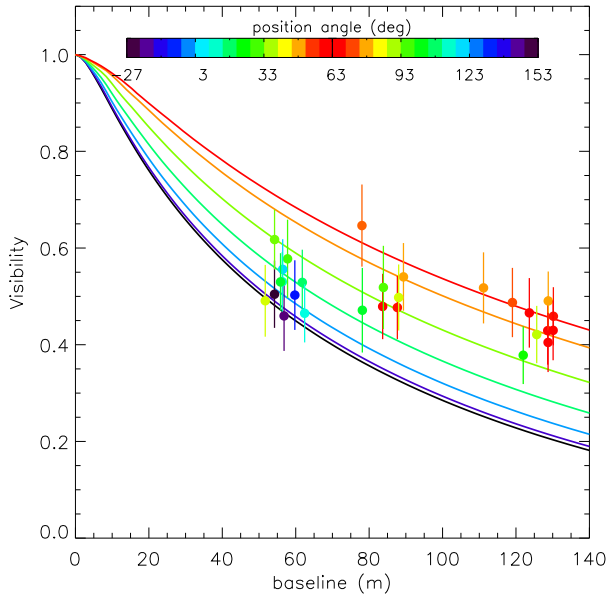}
\caption{11\,$\micron$ visibilities plotted against the projected baseline length for all 2009/2010 observations except the UT1--UT3 data (see note in the text). The position angles for each data point is color coded (from dark-blue along the modeled major axis at $-27^{\prime\prime}$ to red along the minor axis in PA 63$^{\prime\prime}$). Overplotted are PA-dependent visibilities for the $T/d$-gradient model (w/o $L/M$ photometry) as described in Table.~\ref{tab:mod:fit}.}\label{fig:res:visbl}
\end{center}
\end{figure}
 
For illustration we also fitted the model to the photometry-only data. The model using the hot ring + temperature gradient is basically unconstrained with the range of parameters of $2\times\chi^2_{r;\mathrm{min}}$ given in Table~\ref{tab:mod:fit}. The same happens when using the $T/d$-gradient model on the MIDI-only photometry (i.e. the $8-13\,\micron$ total flux spectrum). Both limits are, at best, marginally consistent with the fits involving interferometry. Better constraints are achieved for the $T/d$-gradient model fitted to the mid-IR photometry without the $L$- and $M$-band fluxes. Although the range of parameters allowed is still considerable, it is in overall agreement with the fits including interferometry. This illustrates that the spatial information is important to properly constrain such models. Using complex models on SED-only data may, therefore, lead to results contradictory to the actual spatial distribution of the dust, unless the wavelength range and/or parameters are carefully chosen \citep[for a more detailed discussion with respect to clumpy torus models see][]{Hon10b}.

Finally, we show the 11\,$\micron$ visibilities versus baseline in Fig.~\ref{fig:res:visbl} for all baseline configurations, except UT1--UT3. The data sets for this configuration have been taken under the worst observing conditions, which resulted in difficult calibration, large error bars, and the lowest SNR of all configurations (see Table.~\ref{tab:obs}). Therefore, we exclude these sets from the plot for clarity. The PA for each data point is color coded to be able to distinguish the baseline configurations closer to the major axis (towards dark-blue) and to the minor axis (towards red) as set by the model orientation. As discussed before, the inhomogeneous distribution of telescope position angles and projected baselines make a direct assessment of any kind of elongation difficult. However when concentrating on the shorter baseline lengths (50--60\,m), a trend is seen that the blue data points show systematically lower visibilities than the green points. At the longest baselines ($>$100\,m), the green data points, about in between major and minor axis, are also systematically lower than the red data that are pointing approximately in the direction of the minor axis in the model. As a comparison and further way to prove the elongation and direction of the mid-IR source, we overplotted the visibility curves for different position angles based on the $T/d$-gradient model without $L/M$-band photometry. This illustrates that, in spite of the scatter and the inhomogeneity of the data, the mid-IR source shape and radial brightness distribution are well reproduced by the elliptical $T/d$-gradient model.

\section{Discussion}\label{sec:disc}

\subsection{Mid-IR emission elongated in polar region}

The orientation of the major axis can be put in the context of the AGN system axis in NGC~424. Although \citet{Nag99} report that their 20\,cm/1.5\,GHz radio observations may indicate a marginal extension in about the same direction as their beam size, this could not be confirmed with their higher resolution 3.6\,cm/8.4\,GHz data. \citet{Mun00} consider the source to be unresolved at both frequencies in their data with better SNR, in line with the analysis by \citet{The00}. Consequently, a possible linear radio structure that could indicate the system axis is not detected. 

\citet{Dur87} report [\ion{O}{3}] imaging observations of the nucleus. They note an elongation of the [\ion{O}{3}] in PA$\sim70^\circ$. The extension is seen best in the faintest two contours of 1.25 and 2.5\% relative to their peak contour. The authors note that the determination of the continuum (dominated by the host galaxy) is uncertain, which may result in problems subtracting the continuum emission from the line image. They give an elongation of the [\ion{O}{3}] emission in PA $70^\circ$, with the galaxy major axes in PA $59^\circ$ (i.e. $\Delta$PA = 11$^\circ$). Unfortunately, no HST image at higher quality is available.

\citet{Mor00} report results from optical spectro-polarimetry of the nuclear region of NGC~424. While the unpolarized spectrum shows signs of a strong host component of 45\% (quite typical for Seyfert 2 AGN), the polarized flux displays broad emission lines, which indicates a hidden type 1 nucleus. From their polarization data, the authors give a polarization angle of 43$^\circ$. In type 2 AGN, the polarization angle has been found to be dominated by scattering (either electrons or dust) in the polar region of the AGN and be approximately orthogonal to the linear radio axis where such a feature was detected \citep[e.g.][]{Ant83,Ant93,Ant01,Smi04}. It is, therefore, also orthogonal to the symmetry axis of the torus. In order to confirm that the scatterer is located in the polar region, we estimated the intrinsic polarization of the scattered light. \citet{Mur98} report deep spectral observations of the H$\alpha$ line and detect shallow broad wings to the otherwise narrow emission line. We extracted the flux from their spectrum and compared it to the polarized (Stokes) flux from \citet{Mor00}. The broad component of the H$\alpha$ line shows about 15\% polarization, which is typical for scattering over a sizable solid angle as in the polar region of the AGN \citep[e.g.][]{Ant85}.

Based on the polarization measurements, the system axis for NGC~424 would be established in approximately PA $43^\circ - 90^\circ = -47^\circ$. This is not consistent with the [\ion{O}{3}] PA. Both methods are not accurate, though, and offsets have to be expected \citep[e.g.][]{Ant01,Sch03,Smi04}. In the particular case of NGC~424 -- based on our analysis of the polarization data and given the mentioned uncertainties with the [\ion{O}{3}] data -- we consider the polarization data as the more reliable method to estimate the system axis. As a consequence, we expect that the system axis is close to PA $-47^\circ$, which is very close to the direction of the major axis of the mid-IR emission (PA $-27^\circ$), meaning that the parsec-scale mid-IR emission shows elongation in the polar direction. 

\subsection{Polar elongation in other AGN}

The elongation of the mid-IR emission in the polar direction, instead of the ``disk/torus plane'', has been observed recently in some objects \citep[e.g.][]{Boc00,Pac05}.

\paragraph{Circinus galaxy:} \citet{Pac05} shows mid-IR images of the AGN in the Circinus galaxy at 8.8\,$\micron$ and 18.3\,$\micron$ with some extended emission $\sim$40\,pc to the East and West of the nucleus that coincides spatially with the rim of the known [\ion{O}{3}] outflow cone. \citet{Reu10} compare their VLT/VISIR 11.8 and 18.7\,$\micron$ images to near-IR [\ion{S}{7}] coronal line images and demonstrate that the mid-IR extended emission traces the NLR region. \citet{Pac05} further note that this extended emission is responsible for about 30\% of the total emission in the central 3$\farcs$5 while the other $\sim$70\% originate from the unresolved nucleus ($<$20\,pc in their images). Both the color temperature and intensity of the extended emission are in good agreement with optically-thin dust in local thermal equilibrium and heated by the AGN. 

\paragraph{NGC~1068:} The outflow region of NGC~1068 displays significant mid-IR emission that extends out to more than 50\,pc from the nucleus \citep[e.g.][]{Boc00,Tom01} and associated with emission from dust clouds \citep[e.g.][]{Tom06}. The extended emission contributes about 70\% to the 11.6\,$\micron$ flux in 1$\farcs$2 or 85\,pc aperture around the nucleus \citep{Mas06}. Yet, on parsec scales the emission is dominated by the circumnuclear dust and extends about perpendicular to the large-scale outflow cavity \citep{Jaf04,Rab09}. It is noteworthy, though, that the elongation in outflow direction has also been observed by Speckle imaging in the near-IR ($H$ and $K$ bands) with a $K$-band size of $1.3\,\mathrm{pc}\times2.8\,\mathrm{pc}$ \citep[error $\pm0.3$\,pc;][]{Wit98,Wei04}. While one may argue about peculiarities of NGC~1068, NGC~424 and Circinus indicate that a polar-extended mid-IR emission is not untypical in AGN. 

\paragraph{Other objects:} \citet{Hon10a} presented mid-IR 8\,m-telescope images of IC~5063 and MCG--3--34--64 that also showed mid-IR emission in polar direction. In these cases, however, the extension is observed on scales of $>$100\,pc, while in NGC~424 we are dealing with a factor of 100 smaller distance. In NGC~1386, \citet{Reu10} report extended mid-IR emission in approximately North/South direction. This is at about the same direction as the [\ion{O}{3}] outflow\footnote{We loosely refer to the narrow emission line regions (NLR) as outflows based on the evidence from both statistical analysis of emission lines in object samples and individual sources \citep[e.g.][Fig.~7]{Sto92,Reu03,Mue11,Ant01}}. \citep{Fer00} and radio-emission position angle \citep{Nag99} in this object. Finally, we note that the type 1 AGN NGC~3783 is another candidate with even stronger elongation in polar direction on parsec scales \citep[][H\"onig et al. 2012, in preparation]{Hon10a,Kis11b}.

\subsection{Where is the torus?}

The extended mid-IR emission in polar region in NGC~424 gives rise to the question if it is consistent with the paradigm that emission from the dusty torus dominates the mid-IR emission. This paradigm is well summarized in \citet[their Fig.~4]{Mor09} where the authors show results from decomposition of Spitzer SEDs of type 1 AGN into the torus, a dust NLR, and an additional hot dust components. According to the median of all decompositions, the torus contributes $\ge$80\% of the flux between 7 and 15\,$\micron$ and dominates ($>$50\%) in the range from 4 to 25\,$\micron$ (the long-wavelength side being limited by wavelength coverage of the data). However, our data suggests that in NGC~424 dust in the polar region dominates the mid-IR emission. Indeed, we do see elongation in polar direction at all observed baseline lengths, i.e. a consistency in elongation toward polar direction at these spatial scales. At the longest baseline lengths, the visibilities have dropped to $\sim$0.4, meaning that \textit{at least} 60\% of the mid-IR single-aperture/total flux show this elongation direction. Thus, it is not too far reaching to interpret this as evidence that $\ga60\%$ of the mid-IR total flux emission originates from the polar region.

Interestingly, \citet{Tri12} report on new mid-IR interferometric observations of the Circinus galaxy on short baseline lengths, complementing previous results on longer baselines \citep{Tri07}. These new data indicate that the mid-IR emission is extended in outflow direction not only in the single-telescope aperture as mentioned above, but also on the new baselines. Only at baseline lengths longer than $\sim$50\,m does the orientation of the mid-IR emission (described as a 0.4\,pc ``disk'') change direction toward perpendicular to the outflow region and, therefore, in the direction expected for the classical torus. However, we note that at these baselines the visibilities decreased to $\sim$0.15, meaning that 85\% of the single-aperture mid-IR flux originates from the polar region on scales as small as few parsecs. 

Whether our data for NGC~424 can be considered consistent with the paradigm of mid-IR emission primarily originating from the torus depends on the question if torus models can reproduce the elongation in polar direction in an obscured AGN. It might be possible that self-shielding effects in the mid-plane of the torus cause the torus to \textit{appear} elongated in polar direction while the dust is still strongly concentrated to the mid-plane. An additional factor could be anisotropic illumination from the BBB/accretion disk onto the torus, i.e. radiation escaping preferentially in polar direction. Detailed radiative transfer models that also include visibility studies have been presented by \citet[smooth and clumpy]{Sch08} and \citet[clumpy]{Hon10b}, the former also invoking anisotropy of the accretion disk. Both studies agree that for sets of models that do well represent the IR SEDs as observed for typical AGN, the visibilities should be higher in polar direction (i.e. smaller size) than along the mid-plane of the torus in type-2-like edge-on viewing angles, and more circular towards face-on orientations. Strong polar extension is not to be expected. This also means that although there is self-shielding in the equatorial plane of the torus, the dust mass in typical radiative transfer models is not sufficient to cause strong-enough obscuration that the mid-IR emission is seen elongated in polar direction.

The radiative transfer models use ad-hoc parameterizations of the dust distribution, while hydrodynamical simulations attempt at getting the mass distribution self-consistently from global parameters of the AGN and its environment. \citet{Wad09} present high-resolution numerical simulations of ISM accretion in the inner 32\,pc of an AGN. Although the authors did not perform radiative transfer simulations of the hydrodynamic models, the mass density distribution in the inner parsecs is quite flat (disk-like). At larger radii ($>$10\,pc), the disk flares resulting in a scale hight of the order unity, which is also typically used in the previously mentioned radiative transfer models. Therefore, it is not expected that these models result in strong polar-extended mid-IR emission. \citet{Sch09} present hydrodynamic simulations of the central 100\,pc around an AGN, assuming that the dust is produced in-situ by a stellar cluster. The authors show (log-scale) mid-IR radiative transfer images and SEDs of their standard model. Indeed, these models do show elongation in polar direction and might be considered a better match to our observations. As the authors note, the polar elongation is the result of heavy self-absorbing in a dense torus-plane region, so that only the funnels of their mass distribution are illuminated. As a consequence, the overall IR emission is dominated by cool dust resulting in very red SEDs, incompatible with NGC~424 \citep[note that the SED of the type 2 AGN NGC~424 is significantly bluer in the mid-IR than those of type 1 AGNs NGC~4151 and Mark~841 that have been modeled using these simulations; see Fig. 10 in][]{Sch09}. In particular, the observed silicate emission feature in NGC~424 would require a orientation close to face-on in these models, which, in turn, would make the mid-IR emission appear rather circular (i.e. dominated by the inner disky structure as noted by the authors) and probably inconsistent with the $a/b=2.2$ axis ratio in NGC~424. Additionally, the observed high Hydrogen column density would be difficult to reproduce. 

As mentioned above, self-absorption of mid-IR radiation in the plane of the torus might be a possibility for the polar extension of the mid-IR emission. If the polar-extended mid-IR emission was caused by a high optical depth in the mid-IR (i.e. the mid-IR emission of the dust in the torus gets reabsorbed), then we would have to expect that the ``missing'' mid-IR emission from the plane is re-emitted at longer wavelengths. Consequently we would see either a small bump in the nuclear SED at longer wavelengths (where the optical depth becomes smaller than unity), or, at least, that the nuclear SED appears very red \citep[as illustrated by the hydrodynamic simulations of][]{Sch09}. In NGC~424, however, neither is seen. On the contrary, longward of the MIDI-covered $8-13\,\micron$ range and the 18\,$\micron$ silicate feature, the SED turns down (see Fig.~\ref{fig:irsed})\footnote{We note that the photometric apertures involved in the data reduction do not cause any artificial cut-off at these wavelengths.}, meaning that there is probably no significant amount of radiating dust at longer wavelengths outside of what we see with MIDI. 

\subsection{A possible change of paradigm}\label{sec:para}

Concluding our analysis from the previous section, it is very unlikely that in NGC~424 the elongation in polar direction is (directly) associated with the dusty torus. It seems much more likely that $\ga$60\% of the mid-IR emission is generic to the polar region. When also considering the results by \citet{Tri12} on Circinus, we may argue that the torus becomes energetically insignificant in the mid-IR in these two sources (and probably NGC~3783). Considering that a few other sources also show extension in polar direction in the mid-IR on larger scales (and, in contrast, there are no clear cases with dominant disk-like mid-IR emission), it is not reaching too far to discuss possible implications for AGN in general.

First, we will have to address the nature of the emission in the polar region in NGC~424. As indicated by our model results shown above, the emission is consistent with dust in thermal equilibrium with the AGN. It is, therefore, reasonable to assume that the mid-IR emission is indeed originating from dust. Since we know that $\ga$60\% of the mid-IR single-aperture emission originates from the polar region, we can use the IR SED in this wavelength region as a further constraint. The IRS and MIDI spectra show weak silicate emission features at 10 and 18\,$\micron$, which by its own points toward dust as the origin. Moreover, it indicates that the emitting dusty medium is optically thin to its own thermal radiation \citep[see similarities and small anisotropy of mid-IR emission of type 1 and type 2 AGN][]{Hon11a}. Therefore, if we would look through the polar region onto the AGN (type 1 case), the overall column would still be optically thin. This would be consistent with the unification scheme that requires mostly unobscured views of the central engine along face-on line-of-sights (at least in a statistical sense). However, if the dust is optically thin in the mid-IR, the weakness of the silicate features would argue for a deficiency of silicate grains with respect to graphite grains in ISM dust mixtures, as expected when dust is exposed to the BBB radiation field \citep[e.g.][]{Phi89}.

\begin{figure}
\begin{center}
\epsscale{1.2}
%\epsscale{0.8}
\plotone{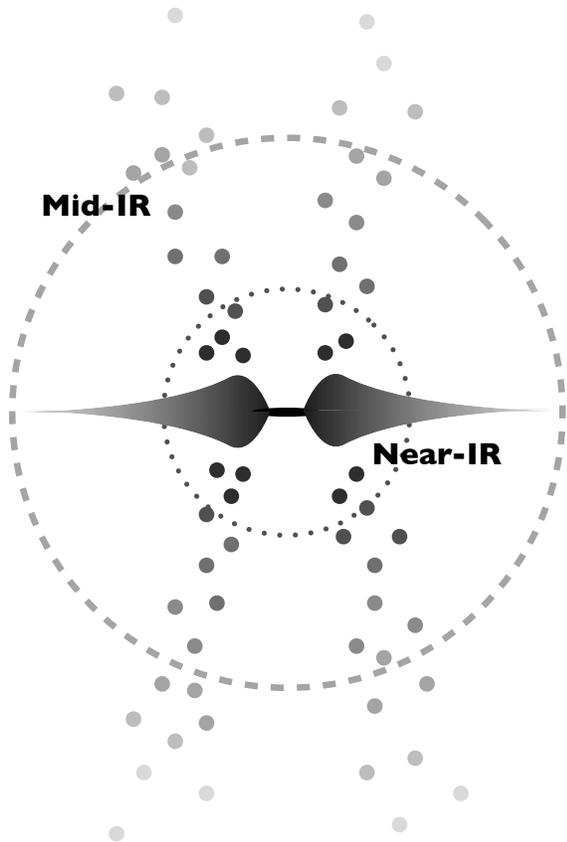}
\caption{Sketch of the dust distribution that contributes to the near- and mid-IR emission around AGN like NGC~424 and Circinus. The classical torus that provides the angle-dependent obscuration is geometrically-thick only in the inner region that radiates primarily in the near-IR. In the mid-IR the emission is dominated by dusty clouds or filaments in polar region that cover a larger surface compared to the torus at these wavelengths. The dashed and dotted annuli indicate the approximate scales from which the near- and mid-IR emission, respectively, originate.}\label{fig:dis:sketch}
\end{center}
\end{figure}
 
NGC~424 is a type 2 AGN and it certainly has an obscuring medium along the line-of-sight to the accretion disk and BLR given the observed high X-ray column density in spite of displaying a silicate emission feature \citep[a mismatch in column density versus optical/IR obscuration is common in AGN;][]{Mai01}. This means that it offers support for the validity of the unification scheme in this source. The unification scheme also requires a geometrically-thick torus -- for which we did not find evidence in the mid-IR here. However, angle-dependent obscuration does not require that the scale height is large at all distances from the AGN, although this is what most torus models assume. It is well possible that the geometrically-thick part that gives rise to the required torus covering factor in the unification scheme is limited to the innermost dusty region. Indeed, as mentioned before, type 1 AGN show a distinct near-IR ``$3-5\,\micron$ bump'' in the SED from hot dust, and this component has different emission characteristics than the mid-IR emission \citet{Mor09,Kis11b}. The size of this component is consistent with AGN-heated, large graphite grains near their sublimation temperature \citep{Kis11a}. Moreover, IR interferometry suggests that this hot-dust component has at least a comparable or even larger surface filling factor as the cool dust \citep[see Fig.~11 of][]{Kis11b}. One interpretation of this result is a considerable covering factor of the hot dust component \citep[see also][for an SED-based argument on the covering factor of the hot component]{Mor09}. We may, therefore, speculate that the classical, obscuring torus is, in fact, related to the hot-dust emitting dust component rather than the mid-IR emitting component.

In Fig.~\ref{fig:dis:sketch} we show a sketch of the dust distribution around NGC~424 as this scenario would suggest. The torus flares strongly in the innermost part, but becomes geometrically thin in the outer regions. In this picture, the near-IR emission originates from the obscuring part of the torus while the mid-IR emission comes primarily from the polar region. The IR-emitting sources in the polar regions have been drawn as clouds, but they could also be filamentary. This inhomogeneous structure of the dust is motivated by the mid-IR interferometry of Circinus \citep{Tri07}, but also mid-IR imaging of NGC~1068 \citep[e.g.][]{Gal05}. Moreover, in Circinus (but not necessarily in NGC~1068), the extended mid-IR emission in the single-telescope images seems to be associated with the edge of the outflow region \citep{Tri07,Tri12}. We also see an offset of 20$^\circ$ between mid-IR position angle and the polarization angle in NGC~424. This could be a sign of an edge-brightening effect as expected in a cone-like region filled with optically-thin dust. Therefore, we placed the polar-region clouds preferentially around the edges of the opening cone in the sketch. Note that this is not a full picture of the distribution of the dust but only a sketch for the IR-emitting material and the polar region may well be filled more homogeneously with dust clouds.

\subsection{Further implications}

This picture also has several implication related to the physical mechanisms around the AGN. The new observations presented here can be interpreted as strong support for a dusty wind \citep[as used in models for the X-ray emission of some AGN; e.g.,][]{Kom97,Kom98,Tur03}. This wind could be driven by radiation pressure on dust grains in the AGN environment. For an AGN in the Seyfert regime radiating at about a tenth of its Eddington luminosity $L_\mathrm{Edd}$ (defined by electron scattering), radiation pressure is sufficient to unbind dust from the gravitational potential \citep[the corresponding dust Eddington luminosity is $L_\mathrm{Edd;dust} \sim 10^{-4\ldots-5} \, L_\mathrm{Edd}$;][]{Pie92,Hon07}. The mid-IR spectra also hint toward the source of the dust that is driven outward in polar direction: as discussed in Sect.~\ref{sec:para} the dust seems to be deficient of silicate grains. Silicates have a lower sublimation temperature than graphite grains \citep{Phi89}, making them disappear from a dusty medium at generally larger distances from the heating source than graphite grains (for a given grain size). This is consistent with the near-IR interferometry and reverberation mapping sizes in AGN that favor large graphite grains for the hottest dust at the sublimation temperature \citep{Kis07,Kis09a}, and similar requirements from optical extinction curves in type 1 AGN \citep{Gas04}. This inner region is also the one closest to the AGN where the dust is directly exposed (i.e. without significant self-obscuration) to the strongest radiation field. It is, therefore, plausible to expect that the dust is driven away from this inner part of the torus. This scenario seems qualitatively in agreement with recent radiation feedback simulations in hydrodynamic simulations \citep{Rot12}. \citet{Kea12} recently presented IR SEDs of wind models that could eventually serve as a starting point for a more quantitative test of dusty winds.

Radiatively-driven winds or radiation pressure in both the optical and IR have also been suggested as a mechanism to support the apparent geometrical thickness of the torus \citep[e.g.][]{Kon94,Elv00,Kro07}. Of course, this mechanism works best in the inner part of the torus close to the sublimation radius because of self-shielding effects further out \citep{Kro07}. As a consequence the geometrical thickness would be limited to just the inner few sublimation radii distances from the AGN where, incidentally, the dust is hot enough to cause the near-IR bump seen in type 1 AGN. As pointed out in \citet{Kis11b}, an analogous near-IR bump in the dusty disks of young stellar objects (YSOs) has been described as a ``puffed-up inner rim'' \citep{Nat01,Dul01}. In this scenario we would see the near-IR bump pronounced or weak/absorbed depending on our line-of-sight to the hotter or cooler side of this compact structure. Unlike in AGN, however, a dusty outflow associated with the puffed-up rim has not been observed. This difference is not unexpected given the much softer radiation field of a star compared to an AGN and the associated lower strength of UV/optical radiation pressure to launch such a wind.

The question remains if objects like NGC~424 and Circinus (and possibly NGC~1068) are a special class of objects or if the dusty polar regions are more generic to the AGN population. As discussed above, there are signs of polar extensions in the mid-IR on scales of 10s of parsecs in other Seyfert galaxies as well, but we do not have the interferometry data to test if this polar component is energetically dominant in these sources. If it is generic, then it will require adjustments to current models of the dust emission of AGN.

\section{Summary and conclusions}\label{sec:summary}

We present VLTI/MIDI mid-IR interferometric observations of the X-ray-obscured type 2 nucleus of NGC~424. The $uv$-coverage allowed us to constrain the position angle-dependent sizes of the mid-IR emission source in the $8-13\,\micron$ wavelength range. We used two different methods to overcome size biases by the inhomogeneous distribution of projected baseline lengths on the $uv$-plane (i.e. beam shape effects). As complementary information, a high spatial resolution IR SED has been compiled. In addition to simple geometric fits, we also modeled photometric and interferometric data simultaneously by a simple model that assumes dust emission in thermal equilibrium with the AGN. This led to the following main results:

\begin{itemize}
\item The mid-IR emission source in NGC~424 is resolved and elongated. Its size depends on position-angle, wavelength, and baseline. For a baseline length of 58\,m we find a size of $(1.7\pm0.2)\,\mathrm{pc} \times (1.2\pm0.2)\,\mathrm{pc}$ at 9.1\,$\micron$, and $(2.2\pm0.4)\,\mathrm{pc} \times (1.9\pm0.4)\,\mathrm{pc}$ at 12.5\,$\micron$. As shown in Fig.~\ref{fig:res:wlsize}, the source size shows a systematic increase with wavelength. A more baseline-independent size can be extracted using the half-light radius $R_{1/2}$ as defined in \citet{Kis11b}. We obtain $R_{1/2} = (2.0\pm0.2)\,\mathrm{pc} \times (1.5\pm0.3)\,\mathrm{pc}$ for the $9-13\,\micron$ wavelength-averaged visibilities of NGC~424.
\item The major axis of the mid-IR emission in NGC~424 points toward PA $-27^\circ\pm2^\circ$ based on the best-fit model of the complete data set. This direction is consistent with the PA derived from simple geometric fits to individual wavelength data corrected for the $uv$-coverage bias (or beam shape). We also detect a flatter radial brightness distribution in the direction of the major axis than along the minor axis of the mid-IR emission. Interestingly, the position angle of the major axis is only 20$^\circ$ off the system axis as set by spectro-polarimetry. Therefore, we conclude that the mid-IR emission is extended in polar direction or symmetry axis of the torus.
\item The polar-elongated emission is responsible for $\ga$60\% of the mid-IR total flux (i.e. single telescope flux) in NGC~424 on scales from about 1\,pc to $\sim$100\,pc.
\end{itemize}

These results can be put into the broader context of AGN unification and our current picture of the mid-IR emission from the dusty torus. We conclude:
\begin{itemize}
\item The elongation of the nuclear mid-IR source on parsec scales in polar direction seems inconsistent with typical state-of-the-art radiative transfer models of the torus, either smooth or clumpy. While the hydrodynamic simulations by \citet{Sch09} do show polar-elongated mid-IR emission, the model SEDs are incompatible with the observed IR SED of NGC~424. We conclude that it is very difficult to simultaneously explain both the shape and SED of NGC~424 within the framework of current torus models.
\item Based on the spectral (SED) and spatial (interferometry) information, the source of the emission is most likely optically thin dust in the polar region of the AGN. While the total dust column and surface brightness are, therefore, low, the emission covers a relatively large area leading to the high contribution of flux to the overall mid-IR emission. We note that similar mid-IR extensions in polar region have been observed on scales of several 10s of parsecs in a few other sources, including NGC~1068.
\item The characteristics of the mid-IR emission in NGC~424 show similarities to the nucleus of the Circinus galaxy (and probably NGC~1068 as well) where about $\sim$85\% of the mid-IR flux originates in the polar region and coincides with the western edge of the outflow cone., i.e. several degrees off from the nominal system axis. In NGC~424, the mid-IR major axis is off about 20$^\circ$ from the polar axis. This could be interpreted as an edge brightening effect in a cone filled with optically thin dust.
\item Our data suggests that the origin of the dust in the polar region may be a radiatively-driven wind from the inner part of the torus. The dust in the inner torus region is expected to be deficient of silicate grains, which is well matched by the mid-IR spectral features of the polar dust. Based on recent near- and mid-IR interferometry results of a sample of type 1 AGN, we propose that the torus has a small scale height at large distances from the AGN and is puffed-up by radiation pressure in the inner region, leading to the required covering factors in the unification scheme. In this picture, the near-IR emission (up to about $5\,\micron$ would be dominated by the torus, while the dusty outflow would be the main contributor to the mid-IR.
\end{itemize}
Our results for NGC~424 have significant implications for models of the IR emission of AGN. Owing to a lack of suitable data, we do not know if the mid-IR emission sources in NGC~424 and a few other nearby Seyfert galaxies are special cases or more generic to AGN, but it seems as more cases of polar-extended mid-IR emission are revealed when objects are studied in detail by high-angular resolution techniques. Moreover, the results of \citet{Hon10a} and \citet{Kis11b} suggest that the structure of the dusty environment depends on luminosity. It is, therefore, important to follow-up our results with interferometry from near- to mid-IR wavelengths and a good position-angle and baseline coverage of more sources. The upcoming VLTI/MATISSE instrument, operating from 3.5 to 13\,$\micron$, should be able to test if the near-IR emission is, indeed, dominated by the torus while the mid-IR flux originates in the outflow. 

\acknowledgments

\begin{footnotesize}
\textbf{Acknowledgements ---} We would like to thank the anonymous referee for the thorough assessment of our manuscript and the comments that helped improving it. Furthermore, we want to thank J. Gallimore for providing the Spitzer data of NGC~424, and E. Moran, P. Gandhi, and D. Asmus for helpful discussions. This research is based on observations made with the European Southern Observatory telescopes under programs 080.B--0332, 083.B--0288, and 086.B--0019. S.F.H. acknowledges support by Deutsche Forschungsgemeinschaft (DFG) in the framework of a research fellowship (“Auslandsstipendium”). This research has made use of the NASA/IPAC Extragalactic Database (NED) which is operated by the Jet Propulsion Laboratory, California Institute of Technology, under contract with the National Aeronautics and Space Administration. Based, in part, on data obtained from the ESO Science Archive Facility.
\end{footnotesize}

\clearpage
\appendix

\section{NGC 424 VLTI/MIDI interferometry data (Online-only)}

In Fig.~\ref{fig:app:intf} we present all reduced and calibrated correlated flux data (blue lines with error bars) and visibilities (gray lines) obtained from 2007 through 2010. The telescope configuration, projected baseline length, position angle, and observing dates are indicated. For more details on the data quality we refer to the maximum SNR values of each set indicated in Table~\ref{tab:obs}. As discussed in Sect.~\ref{mod:data} the shortest wavelength data points are probably affected by decoherencing effects or the wing of the remaining 7.7\,$\micron$ PAH feature. They have been excluded from the model fitting as well as the 2007 data points (see Sect.~\ref{sec:model} for details). The best-fit model visibilities are overplotted as green-dashed lines in Fig.~\ref{fig:app:intf}. Note that some of the data sets show residuals of the atmospheric ozone feature at $9.6\,\micron$ that remained after calibrating the data.

\begin{figure*}[b]
\begin{center}
\epsscale{0.95}
\plotone{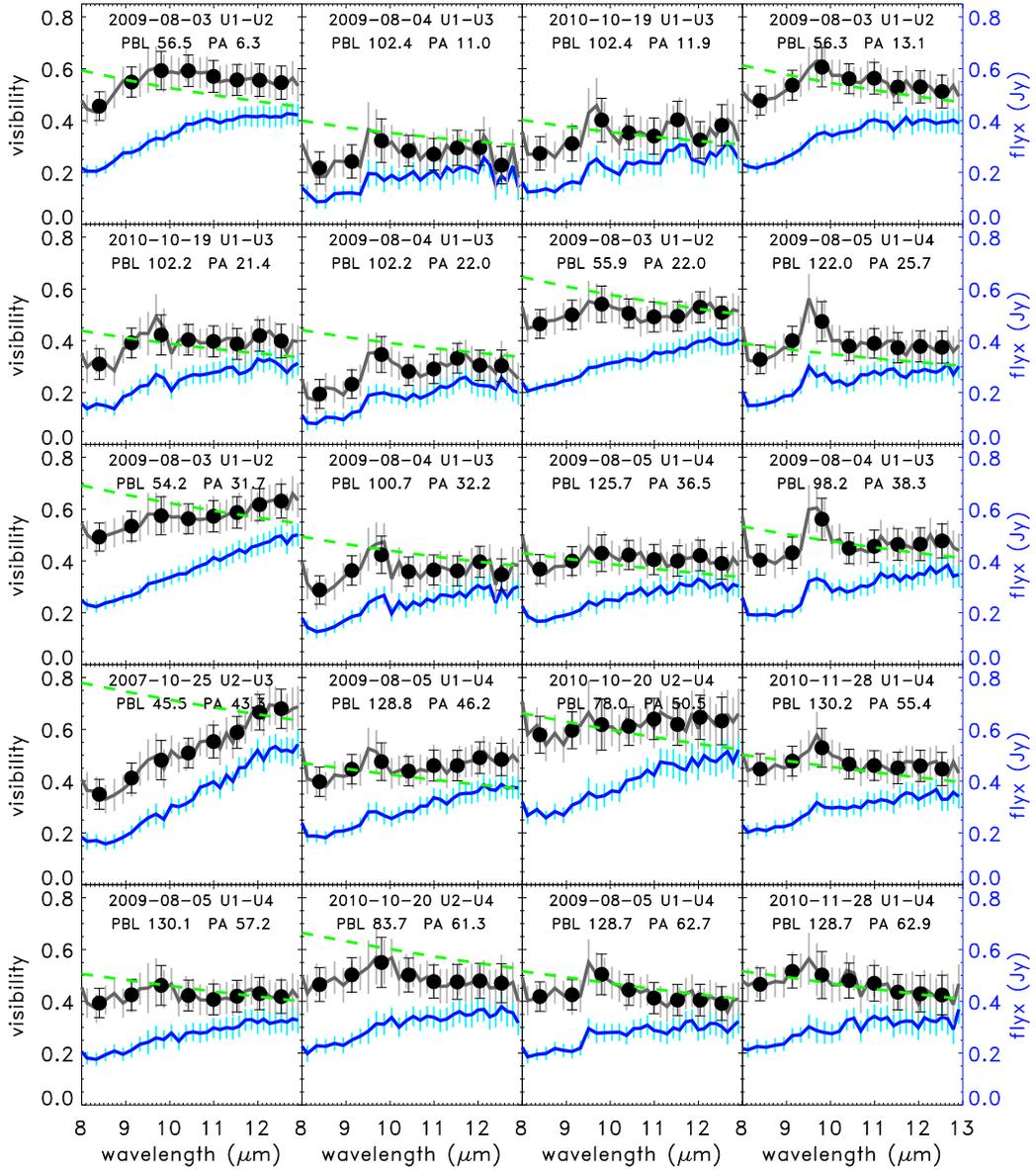}
\caption{All 37 reduced and calibrated interferometric data sets obtained for NGC~424. The blue lines with error bars show the correlated fluxes. The visibilities are shown as gray lines with error bars. We also overplotted the binned visibilities that have been used for size extraction and model fitting (black-filled circles with error bars). The green-dashed line represents the hot ring + $T/d$-gradient model fit to all interferometric and photometric data as detailed in Sect.~\ref{sec:model}.}\label{fig:app:intf}
\end{center}
\end{figure*}

\addtocounter{figure}{-1}
\begin{figure*}
\begin{center}
\epsscale{0.95}
\plotone{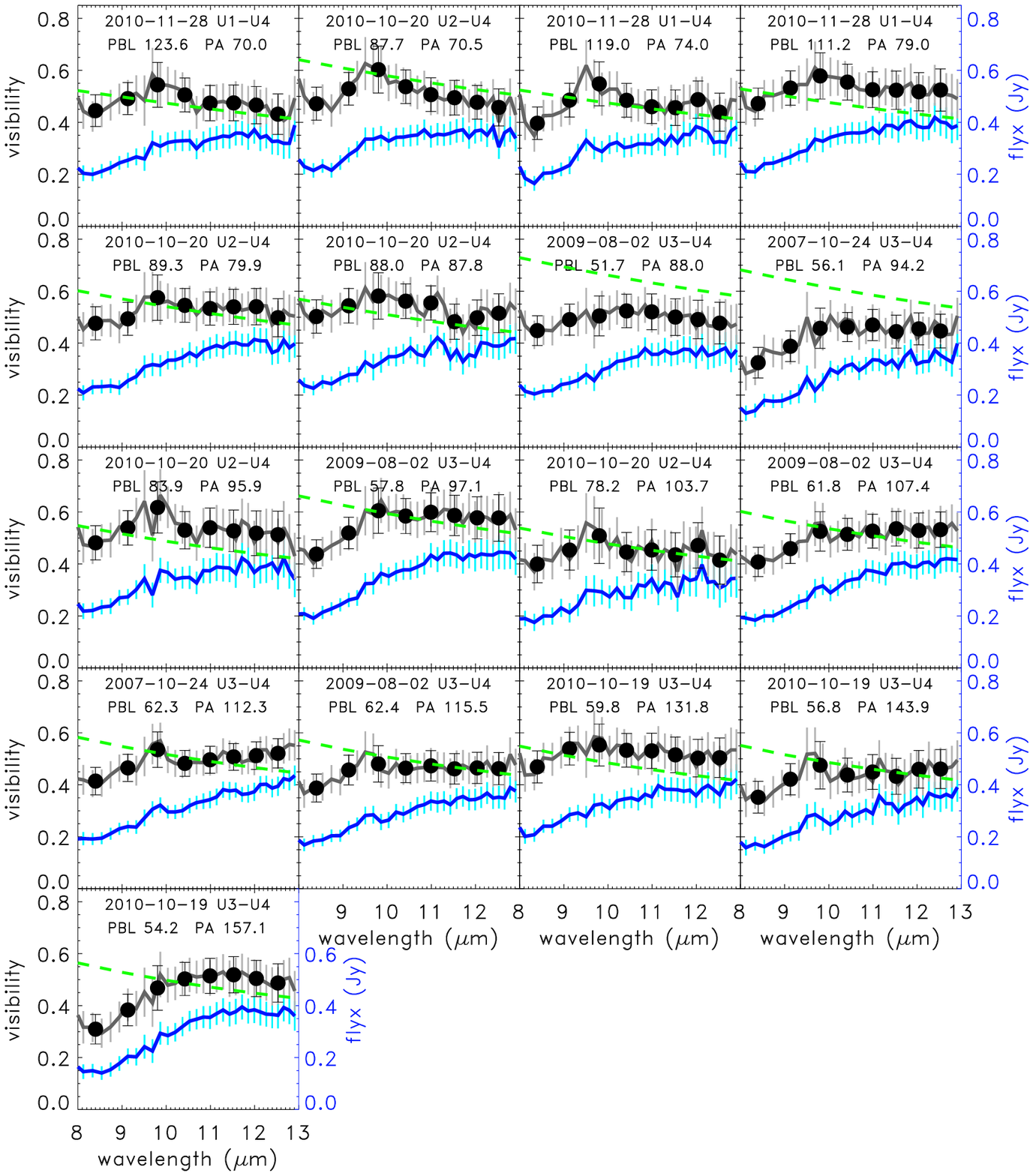}
\caption{continued.}
\end{center}
\end{figure*}

\end{document}